\documentclass[11pt,a4paper]{article}
\pdfoutput=1
\usepackage{jheppub}

\usepackage{multirow, graphicx,amssymb,url,mathrsfs,amsmath}
\usepackage{wrapfig,boxedminipage,setspace,subfigure,epsfig}
\usepackage{amsxtra,amstext,latexsym,dsfont,amsfonts}
\usepackage{color,eucal}

\newcommand{\dd}{\mathrm{d}}

\newcommand{\tg}{{\mathcal{G}}}

\title{A Simple Holographic Superconductor with Momentum Relaxation}

\author[a]{Keun-Young Kim,}
\author[a]{Kyung Kiu Kim,}
\author[b]{and Miok Park}

\emailAdd{fortoe@gmail.com}
\emailAdd{kimkyungkiu@gmail.com}
\emailAdd{miokpark76@gmail.com}

\affiliation[a]{ School of Physics and Chemistry, Gwangju Institute of Science and Technology,
Gwangju 500-712, Korea
}
\affiliation[b]{School of Physics, Korea Institute for Advanced Study, Seoul 130-722, Korea
}

\abstract{We study a holographic superconductor model with momentum relaxation due to massless scalar fields linear to spatial coordinates($\psi_I =  \beta \delta_{Ii} x^i$), where $\beta$ is the strength of momentum relaxation. In addition to the original superconductor induced by the chemical potential($\mu$) at $\beta=0$, there exists a new type of superconductor induced by $\beta$ even at $\mu=0$. It may imply a new `pairing' mechanism of particles and antiparticles interacting with $\beta$, which may be interpreted as `impurity'.  Two parameters $\mu$ and $\beta$ compete in forming a superconducting phase. 
As a result, the critical temperature behaves differently depending on $\beta/\mu$. It decreases when $\beta/\mu$ is small and increases when $\beta/\mu$ is large, which is a novel feature compared to other models.  
After analysing ground states and phase diagrams for various $\beta/\mu$, 
we study optical electric($\sigma$), thermoelectric($\alpha$), and thermal($\bar{\kappa}$) conductivities. 
When the system undergoes a phase transition from a normal to a superconducting phase, $1/\omega$ pole appears in the imaginary part of the electric conductivity, implying infinite DC conductivity. 
If $\beta/\mu <1$, at small $\omega$, a two-fluid model with an imaginary $1/\omega$ pole and the Drude peak works for $\sigma$, $\alpha$, and $\bar{\kappa}$, but  if $\beta/\mu >1$ a non-Drude peak replaces the Drude peak. It is consistent with the coherent/incoherent metal transition in its metal phase.  The Ferrell-Glover-Tinkham (FGT) sum rule is satisfied for all cases even when $\mu=0$.
  }


\begin{document}

\maketitle

\section{Introduction}

Holographic methods (gauge/gravity duality) have provided novel tools to study diverse strongly correlated systems~\cite{CasalderreySolana:2011us,Hartnoll:2009sz,Herzog:2009xv, Iqbal:2011ae}.  In particular, after a pioneering model of superconductor in holographic methods by Hartnoll, Herzog, and Horowitz(HHH) \cite{Hartnoll:2008vx, Hartnoll:2008kx}, there have been extensive development of the model. We refer to \cite{Hartnoll:2009sz, Herzog:2009xv, Horowitz:2010gk}   for reviews and references. 

The HHH model is translationally invariant. 
Because a  translationally invariant system with finite charge density cannot relax momentum,  the HHH model will exhibit an infinite electric DC conductivity even in the normal metal phase. Therefore, to construct more realistic superconductor models, it is important to incorporate momentum relaxation in the framework of holography.

One way to include momentum relaxation is to break translational invariance by imposing explicit inhomogeneous boundary conditions such as a spatially modulated scalar field or temporal $U(1)$ gauge field $A_t$, which mimicks an ionic lattice~\cite{Horowitz:2012ky, Horowitz:2012gs}.  These models successfully yield a finite DC conductivity as well as interesting features in optical conductivity  such as a Drude-like peak at small $\omega$ and some scaling laws 
at intermediate $\omega$. We refer to \cite{Ling:2013nxa,Chesler:2013qla,Donos:2014yya} for further development.   
This idea (in particular, explicit optical lattice) was applied to the HHH model in \cite{Horowitz:2013jaa} and, interestingly, many properties of the bismuth-based cuprates were observed. 

In this method, however, because of inhomogeneity of dynamic fields, the equations of motion become complicated coupled partial differential equations(PDE). It is technically involved and less flexible than ordinary differential equations(ODE), though conceptually clear. Therefore, it will be efficient and complementary if we can analyse 
the system with ODEs.  In this line a few ideas have been proposed and developed. 

Massive gravity models~\cite{Vegh:2013sk,Davison:2013jba,Blake:2013bqa,Blake:2013owa} introduce mass terms for some gravitons. It breaks bulk diffeomorphism invariance and consequently breaks translation invariance in the boundary field theory. Holographic Q-lattice models~\cite{Donos:2013eha,Donos:2014uba} exploit a continuous global symmetry of the bulk theory, where, for example, the global phase of a complex scalar field breaks translational invariance. Models with massless scalar fields linear in spatial coordinate~\cite{Andrade:2013gsa,Gouteraux:2014hca, Taylor:2014tka, Kim:2014bza} take advantage of the shift symmetry\footnote{This model with analytic solutions have been reported in \cite{Bardoux:2012aw} without specific applications to gauge/gravity duality.   In holographic context a model with only one scalar field was studied for an anisotopic background in \cite{Iizuka:2012wt, Cheng:2014qia}}. This model is related to Q-lattice models. For example, a massless complex scalar with constant $\varphi$ in (2.6) of \cite{Donos:2013eha} gives a massless axion linear in a spatial direction. Also there are models  utilising a Bianchi VII$_0$ symmetry to construct black holes dual to helical lattices~\cite{Donos:2012js, Donos:2014oha,Donos:2014gya}. All these models give us ODEs and yield a finite DC conductivity as expected. Furthermore, for a large class of models, the analytic DC conductivity formulas are available in terms of the data of the black hole horizon~\cite{Donos:2014cya}.

Building on this development based on ODE, it is natural to revisit the holographic superconductor models.
The superconductor model combined to the massive gravity models and Q-lattice models have been studied in \cite{Zeng:2014uoa} and  \cite{Ling:2014laa, Andrade:2014xca} respectively. For massless scalar models, 
an anisotropic background case with one scalar field was addressed in \cite{Koga:2014hwa, Bai:2014poa}. 
As in non-superconducting cases, the properties of theses ODE-based superconductor models 
qualitatively agree to the PDE-based model with ionic lattice~\cite{Horowitz:2013jaa}.

In this paper, we study a holographic superconductor model based on a massless scalar model for isotropic background. 
The model consists of two parts: The HHH action~\cite{Horowitz:2013jaa} and two massless real scalar fields~\cite{Andrade:2013gsa}. The HHH action is a class of Einstein-Maxwell-complex scalar action with negative cosmological constant. 
An on shell massless real scalar field($\psi$) is linear to spatial coordinate with proportionality constant $\beta$, for example, $\psi  =  \beta x$.  To have isotropic bulk fields, the identical scalar field is  introduced for every spatial direction in field theory so there is only one parameter, $\beta$, which controls the strength of momentum relaxation.

The HHH model without massless real scalar sector has been studied extensively. See ~\cite{Hartnoll:2009sz,Herzog:2009xv} for review. It has two phases, normal metallic phase and superconducting phase.  Without the complex scalar field, the black hole  is Reissner-Nordstrom type and the system is in normal metal phase. With a finite complex scalar hair, the system is in  superconductor phase. With massless real scalar fields, a normal metal phase still exist~\cite{Bardoux:2012aw,Andrade:2013gsa} and its thermodynamic and transport coefficients were studied: the DC electric conductivity~\cite{Andrade:2013gsa}, DC thermoelectric and thermal conductivity~\cite{Donos:2014cya}, optical electric conductivity~ \cite{Taylor:2014tka},  optical  electric, thermoelectric and thermal conductivities~\cite{Kim:2014bza}. 

Having studied the metal phase of the model, we want to investigate the superconducting phase. First, we will examine the condition in which a superconductor phase may exist at finite $\beta$.  Second, we will study the effect of $\beta$ on superconducting phase transition and the properties of superconductor. 

Interestingly, we find that there exists a new type of superconducting phase even when $\mu=0$  at finite $\beta$ in addition to the original superconductivity at $\beta=0$ and finite $\mu$.   i.e. large $\beta>T$ induces superconductivity as large $\mu>T$ does. Because two parameters $\mu$ and $\beta$ compete each other in forming superconducting phase the critical temperature may behave differently for $\beta/\mu  >1$ and for $\beta/\mu  <1$.
At zero temperature the $\beta$ suppresses the superconductivity. At finite temperature for small $\beta/\mu$  the critical temperature decreases, while for large $\beta/\mu$, the critical temperature increases. 
It is a novel feature compared to other models.

Our main tool to analyse the superconducting phase is optical conductivities: electric($\sigma$), thermoelectric($\alpha$), and thermal($\bar{\kappa}$) conductivities. 
In holographic framework, the computation of these conductivities are 
related to the classical dynamics of three coupled bulk field fluctuations(metric, gauge, scalar fields). 
By computing the on-shell quadratic action for these fluctuations we can read off  
the retarded Green's functions relevant to three conductivities.  Most papers deal with only electric optical conductivity.  
However, for better understanding,  it will be good to have a complete set of three conductivities: $\sigma$, $\alpha$, and $\bar{\kappa}$. For a class of models, analytic formulas for DC conductivities are available~\cite{Donos:2014cya, Amoretti:2014mma}, but not for optical conductivities. 
In \cite{Kim:2014bza}, a systematic numerical method to compute all three conductivities 
in a system with a constraint were developed based on  \cite{Amado:2009ts,Kaminski:2009dh}. 
The method was applied to the normal metallic phase of our model, producing numerical conductivities reliably~ \cite{Kim:2014bza}, and we will use the same method for the superconducting phase in this paper\footnote{For another numerical analysis on three optical conductivities we refer to \cite{Amoretti:2014zha}.}. 

One of the main results of  \cite{Kim:2014bza} is numerical demonstrations of coherent/incoherent metal transition\footnote{It was shown in \cite{Cheng:2014tya} for anisotropic background. See also \cite{Davison:2014lua}.}. In \cite{Hartnoll:2014lpa} metal without quasi-particle at strong coupling was classified by two classes: coherent metal with a Drude peak and incoherent metal without a Drude peak.  At small $\omega$, when momentum dissipation is weak ($\beta < \mu$), all three optical conductivities fit well to the Drude form modified by $K_0$:
\begin{equation} \label{DrudeIntro1}
\frac{K_n \tau}{1 - i \omega \tau} + K_0 \,,
\end{equation}
where $K_0$ is the contribution  from pair production affected by net charge density. For $\beta \ll \mu$, $K_0$ can be ignored and a modified Drude form is reduced to the standard Drude from.  If $\beta > \mu$ optical conductivities do not fit to \eqref{DrudeIntro1} and goes to the incoherent metal phase\footnote{See \cite{Amoretti:2014ola} for discussions on  universal bounds for thermoelectric diffusion constants in incoherent metal. }, which agrees to \cite{Hartnoll:2014lpa}. 
In superconducting phase, we find a similar result. If $\beta < \mu$ all three optical conductivities  fit to 
\begin{equation} \label{DrudeIntro2}
i\frac{K_s}{\omega} + \frac{K_n \tau}{1 - i \omega \tau} + K_0  \,,
\end{equation}
where $K_s$ is supposed to be proportional to superfluid density. In superconducting phase, it turns out that $K_s \ne 0$ for $\sigma$ and $\kappa$, but $K_s = 0$ for $\alpha$. 

We have confirmed numerically the Ferrell-Glover-Tinkham (FGT) sum rule is satisfied in all cases we considered. 
\begin{equation} 
\int^{\infty}_{0^+} \dd \omega \mathrm{Re} [\sigma_n (\omega) - \sigma_s (\omega)] = \frac{\pi}{2} K_s  \,,
\end{equation}
where $\sigma_s$($\sigma_n$) is the electric conductivity at $T<T_c$($T>T_c$).

This paper is organised as follows.
In section \ref{sec2},  we introduce our holographic superconductor model (Einstein-Maxwell-complex scalar action with negative cosmological constant) incorporating momentum relaxation by massless real scalar fields. Background bulk solutions corresponding to superconducting phase and normal phase are obtained. By comparing on-shell actions of both solutions, we identify the phase transition temperature as a function of chemical potential and momentum relaxation parameter, which yields 3-dimensional phase diagrams. In superconducting phase, we also compute condensates as a function of temperature for given chemical potential and momentum relaxation parameter. 
In section \ref{sec3}, we compute optical electric, thermoelectric, and thermal conductivities in superconducting phase and normal phase.  In particular, in superconducting phase, we discuss the effect of momentum relaxation on conductivity in several aspects such as the appearance of infinite DC conductivity, Drude-nature of optical conductivity in small frequency range, two-fluid model, and Ferrell-Glover-Tinkham(FGT) sum rule. We also present a general numerical method to compute retarded Green's functions when many fields are coupled.
In section \ref{sec4} we conclude.

{\bf{Note added:}} After this paper was completed, we became aware of \cite{Andrade:2014xca} which has overlap with ours.

\section{Metal/superconductor phase transition}\label{sec2}

We start with  the original holographic superconductor model proposed by Hartnoll, Herzog, and Horowitz(HHH) \cite{Hartnoll:2008vx} 
\begin{align} \label{HHH}
S_{\mathrm{HHH}} &= \int_{M}  \dd^{d+1}x \sqrt{-g} \left[   R + \frac{d(d-1)}{L^2} -\frac{1}{4}F^2   -  |D\Phi|^2 - m^2  |\Phi| ^2   \right] \,,  \\ \label{GH}
S_{\mathrm{GH}} &= - 2 \int_{\partial M} \dd^d x   \sqrt{-\gamma} K ~ ,
\end{align}
where $F= \dd A$ is the field strength for a $U(1)$ gauge field $A$ and $\Phi$ is a complex scalar field.   We have chosen units such that the gravitational constant $16 \pi G = 1$. The second action, $S_{\mathrm{GH}}$, is the Gibbons-Hawking term, which is required for a well defined variational problem with Dirichlet boundary conditions. $\gamma$ is the determinant of the induced metric $\gamma_{\mu\nu}$ at the boundary, and $K$ denotes the trace of the extrinsic curvature.  To impose a momentum relaxation effect, we add the action of free massless scalars proposed in \cite{Andrade:2013gsa} 
\begin{equation} \label{psi}
S_\psi = \int_M \dd^{d+1}x \sqrt{-g} \left[  - \frac{1}{2}\sum_{I=1}^{d-1} (\partial\psi_I)^2  \right] \,.
\end{equation}

The action $S_{\mathrm{HHH}} + S_\psi $ yields the equations of motion for matters\footnote{Index convention: $M,N,\cdots = 0,1,2,r$, and $\mu,\nu,\cdots = 0,1,2$, and $i,j,\cdots = 1,2$.}
\begin{align}
&\nabla_M  F^{MN} +i q ( \Phi^* D^N \Phi - \Phi D^N \Phi^*   )=0 \,, \label{eom1} \\
&\left(D^2 - m^2 \right) \Phi =0 \,, \label{eom2} \\
&\nabla^2 \psi_I =0 \,, \label{eom3} 
\end{align}
where the covariant derivative is defined by $ D_M \Phi=  \left(  \nabla_M -i q A_M   \right)  \Phi $, and the Einstein's equation
\begin{equation} \label{eom4}
\begin{split} 
R_{MN} &- \frac{1}{2}g_{MN} \left(  R + \frac{d(d-1)}{L^2} -\frac{1}{4}F^2   -  |D\Phi|^2 - m^2  |\Phi| ^2  - \frac{1}{2}\sum_{I=1}^{d-1} (\partial\psi_I)^2 \right) \\&= \frac{1}{2} \partial_M \psi_I \partial_N \psi_I + \frac{1}{2} F_{MQ}{F_N}^Q + \frac{1}{2}\left(  D_M \Phi D_N \Phi^* + D_N \Phi D_M \Phi^* \right) \,,
\end{split}
\end{equation}

Since we are mainly interested in  $2+1$ dimensional systems, we will set $d=3$ from now on. In order to construct a plane($x,y$)-symmetric superconducting background, we take the following ansatz,
\begin{equation} \label{ansatz1}
\begin{split}
&\dd s^2 = - \tg(r) e^{-\chi(r)} \dd  t^2 + \frac{\dd r^2}{\tg(r)} + \frac{r^2}{L^2}(\dd x^2 + \dd y^2)~, \\
&A = A_{ t}(r) \dd  t\,,\qquad \Phi = \Phi(r)\,, \qquad \psi_I =  \beta_{Ii} x^i =  \frac{\beta}{L^2} \delta_{Ii} x^i\,,
\end{split}
\end{equation}
where non-zero $A_t(r)$ is introduced for a finite chemical potential or charge density  and non-zero $\Phi(r)$ will be 
a hair of black hole, yielding a finite superconducting order parameter. A special form of $\psi_I$  is included for momentum relaxation, where $\beta$ may be interpreted as a strength of impurity.

Plugging the ansatz \eqref{ansatz1} into the equations of motion \eqref{eom1}-\eqref{eom4},  we have four equations \eqref{eom5}-\eqref{eom8}.
Maxwell's equation \eqref{eom1} yields
\begin{equation}
A_{t}'' + \left(\frac{\chi'}{2} +\frac{2}{r}  \right) A_{t}' - \frac{2 q^2  \Phi^2}{\tg} A_{t} = 0 \,, \label{eom5}
\end{equation}
where $\Phi$ can be taken to be real, since $r$ component of Maxwell's equation implies that the phase of $\Phi$ is constant. The complex scalar field equation  \eqref{eom2}  becomes
\begin{equation}
\Phi'' + \left( \frac{\tg'}{\tg}-\frac{\chi'}{2} + \frac{2}{r}  \right)\Phi' + \left(\frac{q^2 e^{\chi} A_{t}^2}{\tg^2}-\frac{m^2}{\tg} \right) \Phi= 0 \,. \label{eom6}
\end{equation}
The massless real scalar equation \eqref{eom3} is satisfied by the ansatz \eqref{ansatz1}. 
The $tt$ and $rr$ components of Einstein's equations  \eqref{eom4}  give
\begin{align}
&\chi' + r \Phi'^2 + \frac{r q^2  A_{  t}^2 \Phi^2 e^{\chi}}{\tg^2}  = 0 \,,  \label{eom7} \\
&\Phi'^2  + \frac{e^{\chi} A_{{  t}}'^2}{2 \tg} + \frac{2\tg'}{\tg r} +  \frac{2}{r^2} - \frac{6}{\tg  L^2}  + \frac{m^2 \Phi^2}{\tg} +  \frac{q^2  A_{{t}}^2 \Phi^2 e^{\chi}}{\tg^2}   =\frac{- \beta^2}{r^2  \tg L^2} \,.\label{eom8}
\end{align}

We will  numerically solve a set of coupled equations of two second order differential equations(\ref{eom5}-\ref{eom6}) and two first order differential equations(\ref{eom7}-\ref{eom8}) by integrating out from the horizon($r_h$), which is defined by $\tg(r_h) = 0$, to infinity.  It turns out that, by the regularity condition at the horizon, six initial conditions are determined by three initial values:
\begin{equation}
\Phi (r_h)\,, \quad {{A}}'_{t}(r_h)\,, \quad  \chi(r_h) \,,
\end{equation}
with $A_t(r_h) = 0$, for a given $r_h$, $\beta$, $m^2$, and $L$.  From regularity of the Euclidean on-shell action the Hawking temperature($T_H$) is given by
\begin{align} 
T_H &= \frac{\tg'(r_h)e^{-\frac{1}{2} \chi(r_h) }  }{4\pi} \label{THH0} \\ 
&= \frac{r_h}{16 \pi L^2} \bigg(  12 - 2 m^2 L^2\Phi (r_h)^2 -2 L^2  \frac{\beta^2}{2 r_h^2}  -  L^2 e^{\chi(r_h)} (A_t'(r_h))^2   \bigg) e^{-\frac{1}{2} \chi(r_h)  } \,.  \label{THH}
\end{align}

Near boundary($r \rightarrow \infty$) the equations \eqref{eom5}-\eqref{eom8} are solved by the following series expansion. 
\begin{align}
\chi(r) &\sim  \chi^{(0)}   +\frac{(\Phi ^{(1)})^2}{2 r^2}+\frac{4 \Phi ^{(1)} \Phi ^{(2)}}{3 r^3}  \cdots  \,, \label{chi1} \\
\tg(r)&\sim \frac{r^2}{L^2} +\frac{(\Phi^{(1)})^2}{2 L^2} - \frac{\beta^2 }{2 L^2}+  \frac{\tg^{(1)}   }{r L^2} +\cdots \,, \\
A_{t}(r) &\sim  A_t^{(0)} - \frac{ A_t^{(1)}}{r}+ \cdots\,, \\
\Phi(r)&\sim \frac{\Phi^{(1)}}{r } + \frac{\Phi^{(2)}}{r^2}+\cdots \,, \label{Phi1}
\end{align}
where we considered a case with $m^2 = -2/L^2$ to be concrete. The mass $m^2$ is related to the conformal dimension $\Delta$ of an operator in the dual field theory, $m^2 = \frac{\Delta (\Delta - d)}{L^2}$, and the scalar field falls off as $r^{-\Delta}$. For $m^2 = -2/L^2$ and $d=3$, $\Delta=1$ or $2$ as shown in \eqref{Phi1}.

The coefficients of \eqref{chi1}-\eqref{Phi1} are identified with the field theory quantities as follows. 
\begin{equation} \label{coeff1}
\begin{split}
& {A}_t^{(0)} \sim  \mu  \,, \qquad    {A}_t^{(1)}  \sim \rho  \,, \qquad \tg^{(1)}   \sim -\epsilon /2 \,, \\
& \Phi^{(1)} = \mathcal J^{(2)} \,,    \Phi^{(2)} \sim  \langle\mathcal{O}^{(2)}\rangle, \quad  \mathrm{or} \quad  \Phi^{(2)}\sim \mathcal J^{(1)}\,, \Phi^{(1)} \sim  \langle \mathcal{O}^{(1)}\rangle  \,,
\end{split}
\end{equation}
where $\mu$, $\rho$, $\epsilon$, $\mathcal O^{(i)}$  and $\mathcal J^{(i)}$   are  chemical potential,  charge density, energy density, $\Delta =i$ operator  and its source, respectively. So $\mathcal{J}^{(i)}$ should vanish for condensation of $\mathcal{O}^{(i)}$.  See appendix  \ref{apa}  for details.  
$\chi^{(0)} $ should be set to be zero to identify the Hawking temperature of the black hole with the temperature of boundary field theory. Equivalently, we may rescale the time 
\begin{equation}
t \rightarrow a t\,, \quad e^\chi \rightarrow a^2 e^\chi \,,  \quad A_t \rightarrow A_t/a \,,
\end{equation}
where $a = e^{-\chi^{(0)}/2}$, to set $\chi^{(0)} = 0$. In practical computation, this rescaling method is much easier since we have to shoot out from horizon. Then the field theory temperature is computed as
\begin{equation}
T =  e^{\chi^{(0)}/2} T_H  \,,
\end{equation}
where $T_H$ is defined in \eqref{THH}. 
However, analytic formulas from here are presented by assuming $\chi^{(0)} = 0$.

There are two scaling symmetries of the equations of motion.
The first symmetry is
\begin{equation} \label{scaling1}
r  \rightarrow  a_1 r \,, \quad (t,x,y)  \rightarrow a_1 (t,x,y)\,,    \quad L \rightarrow a_1 L \,, \quad q \rightarrow q/a_1 \,, \quad m \rightarrow  m/a_1 \,, \quad \beta \rightarrow  a_1 \beta \,,
 \end{equation}
 and the second one is
\begin{equation}  \label{scaling2}
r  \rightarrow  a_2 r \,, \quad (t,x,y)  \rightarrow (t,x,y)/a_2\,,  \quad A_\mu \rightarrow  a_2 A_\mu  \,, \quad \mathcal{G} \rightarrow  a_2^2 \mathcal{G}\,, \quad \beta \rightarrow a_2 \beta  \,.
 \end{equation}
By taking $a_1=1/L$ and $a_2 = 1/r_h$ we may define new scaled variables with tilde. 
 \begin{equation}  \label{scaling3}
\tilde{r} = \frac{r}{r_h} \,,\quad (\tilde{t}, \tilde{x},\tilde{y})  = \frac{r_h}{L^2} (t,x,y)\,, \quad \tilde{A}_\mu = \frac{L}{r_h} A_\mu\,,  \quad\tilde{m} = m L\,, \quad  \tilde{\mathcal{G}} = \frac{L^2}{r_h^2} \mathcal{G}\,, \quad \tilde{\beta} = \frac{\beta}{r_h} \,.
 \end{equation}
While performing numerics we work in terms of these tilde-variables. In practice, it is equivlalent to set $L=r_H=1$ in numerical computation. However, when we interpret final results, we need to scale them back carefully.

Near horizon we have seven parameters, $\Phi (r_h)$,  ${{A}}'_{t}(r_h)$, $\chi(r_h)$, $r_h$, $\beta$, $m^2$, and $L$.    Taking advantage of three scaling symmetries \eqref{scaling1}-\eqref{scaling3}, we may set $r_h=L=\chi(r_h)=1$. 
We shoot for a given $m^2$ and $\beta$, dialing $\tilde{\Phi} (1)$,  $\tilde{{A}}'_{t}(1)$. 
We shoot out from horizon $\tilde{r}_H=1$ targetting $\tilde{\Phi}^{(1)}=0$ or $\tilde{\Phi}^{(2)}=0$ at $r=\infty$. Thus, the solution is given by a line in a two dimensional configuration space of $\tilde{\Phi} (1)$ and $\tilde{A}'_{t}(1)$.

\subsection{Normal (metallic) phase}

Here we minimally summarize the properties of the metal phase to set up the stage for our paper, referring to \cite{Andrade:2013gsa} for more details. 

The normal phase of the system  corresponds to the solution without condensate 
\begin{equation} \label{normal1}
\Phi=0 \,, 
\end{equation}\
and the analytic solution is given by
\begin{align}
\dd s^2 &= -  \tg(r) \dd t^2 +  \frac{\dd r^2}{\tg(r)}  + \frac{r^2}{L^2}(\dd x^2 + \dd y^2)\,,  \quad \chi(r) = 0 \,, \label{normal2} \\ 
& \qquad  \tg(r) = \frac{1}{L^2}\left(  r^2 -\frac{\beta^2 }{2} -\frac{m_0}{r} + \frac{\mu^2 }{4} \frac{r_h^{2}}{r^{2}}   \right)  \,, \\
A&= \frac{\mu}{L} \left(  1- \frac{r_h}{r}   \right)\dd t    \,,  \label{normal3} \\
\psi_I &=  \beta_{Ii} x^i =  \frac{\beta}{L^2} \delta_{Ii} x^i\,, \label{normal4}
\end{align}
where
$m_0$ is determined by the condition $\tg(r_h) =0$:
\begin{equation}
m_0 = r_h^3 \left(  1+\frac{\mu^2}{4 r_h^2} - \frac{\beta^2}{2 r_h^2}     \right) \,.
\end{equation}
The normal phase of the system is described by a charged black brane solution with non-vanishing $\beta$,
 which describes a metal state with a finite DC conductivity. 
The temperature is given by the Hawking temperature \eqref{THH0} :
\begin{align}
T_H = \frac{\mathcal{G}'(r_h)}{4\pi} = \frac{1}{4\pi L^2} \left( 3 r_h - \frac{\mu^2 +2 \beta^2}{4r_h}    \right) ~~.
\end{align}

There is an important property of the geometry which we will rely on when discussing the instability of the metal phase. 
In the zero temperature limit, the near horizon geometry of an extremal  black brane becomes AdS$_2\times \mathbb{R}^ {d-1}$ with 
the effective radius of AdS$_2$ given by
\begin{align} \label{AdS2}
L_{2}^2 = \frac{L_{d+1}^2}{d(d-1)} \frac{ (d-1) \beta^2 + (d-2)^2 \mu^2 }{  \beta^2 + (d-2)^2 \mu^2  }~~.
\end{align}
Notice that even at $\mu=0$ case, the near horizon geometry remains to be AdS$_2\times \mathbb{R}^ {d-1}$  due to a finite $\beta$. In this sense, $\mu$ and $\beta$ will play a similar role as far as instability is concerned.

\subsection{Criterion for instability  and  quantum phase transitions}

Before performing a full numerical analysis to search superconducting phase, we first want to address a simpler question: When can the normal phase( \eqref{normal1}-\eqref{normal4}) be unstable by small scalar field perturbations, which may develop into a hairy black hole with nonzero $\Phi$? To address this question we perform the analysis presented in \cite{Denef:2009tp,Hartnoll:2009sz}.

We look for an unstable mode of $\Phi = \phi(r) e^{-i\omega t}$ of the equation \eqref{eom2} 
in the background  \eqref{normal1}-\eqref{normal4}:
\begin{align} \label{BF05}
0&=(\nabla_\mu - i q A_\mu )(\nabla^\mu - i q A^\mu )\Phi - m^2 \Phi  \\
&=\mathcal{G}  \left(  \phi''  + \left(  \frac{2}{r} +  \frac{\mathcal{G}'}{\mathcal{G}} \right) \phi'  +  \frac{ \left( q \mu (r-r_h) + r L \omega \right)^2 }{L^2 r^2 \mathcal{G}^2} \phi - \frac{m^2}{\mathcal{G}} \phi   \right) e^{-i \omega t}   \label{BF1}
\end{align}
The normal phase is unstable if there is a normalisable solution with incoming boundary conditions 
at the horizon such that $\omega$ has a positive imaginary part. Because we are interested 
in determining the critical temperature it is enough to search a static($\omega = 0$) normalisable mode.

\begin{figure}[]
\centering
  \subfigure[Transition temperatures at $\beta/\mu = 1$]
   {\includegraphics[width=6cm]{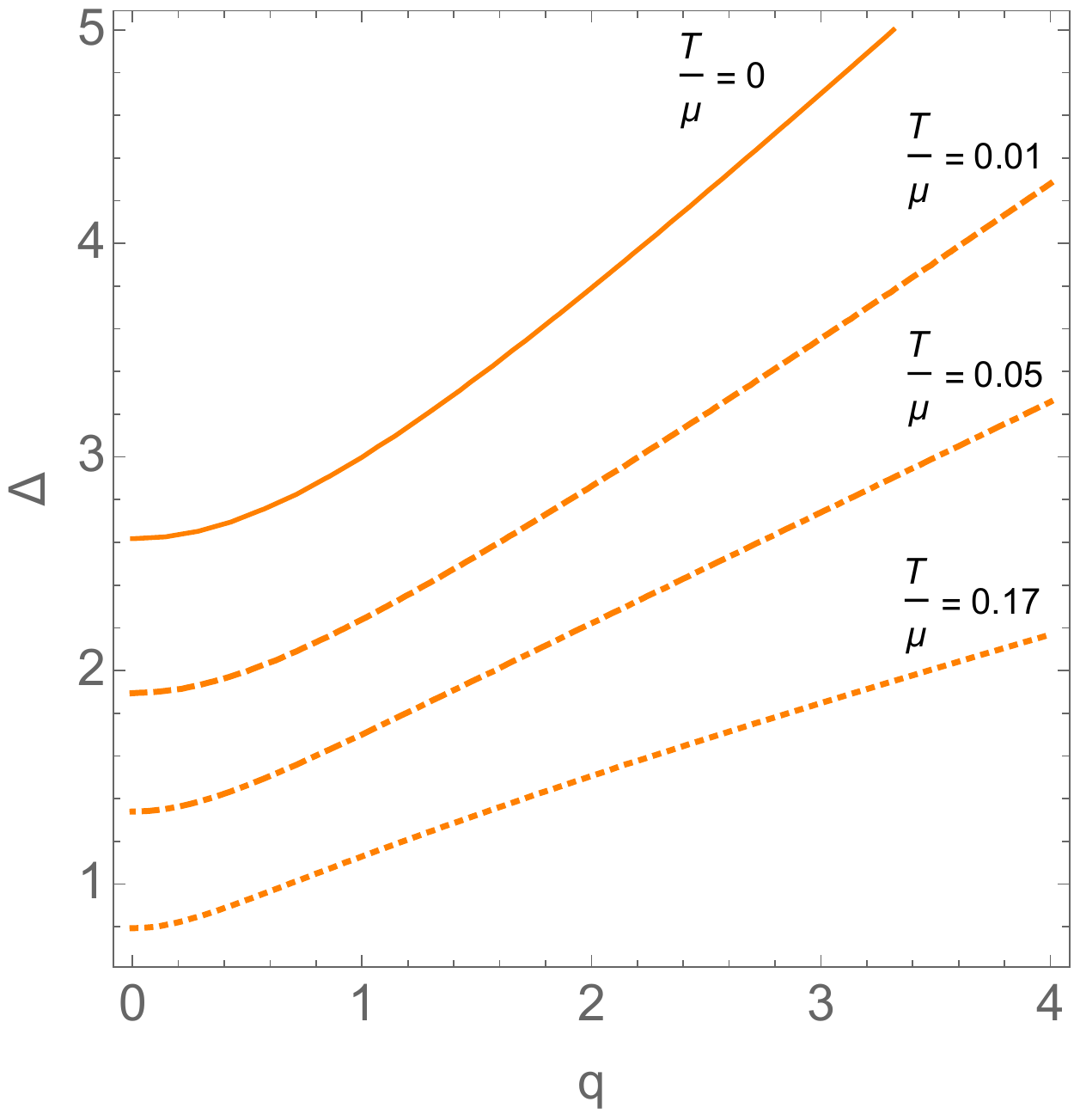} \label{}} \ \
     \subfigure[Quantum phase transitions ]
   {\includegraphics[width=6cm]{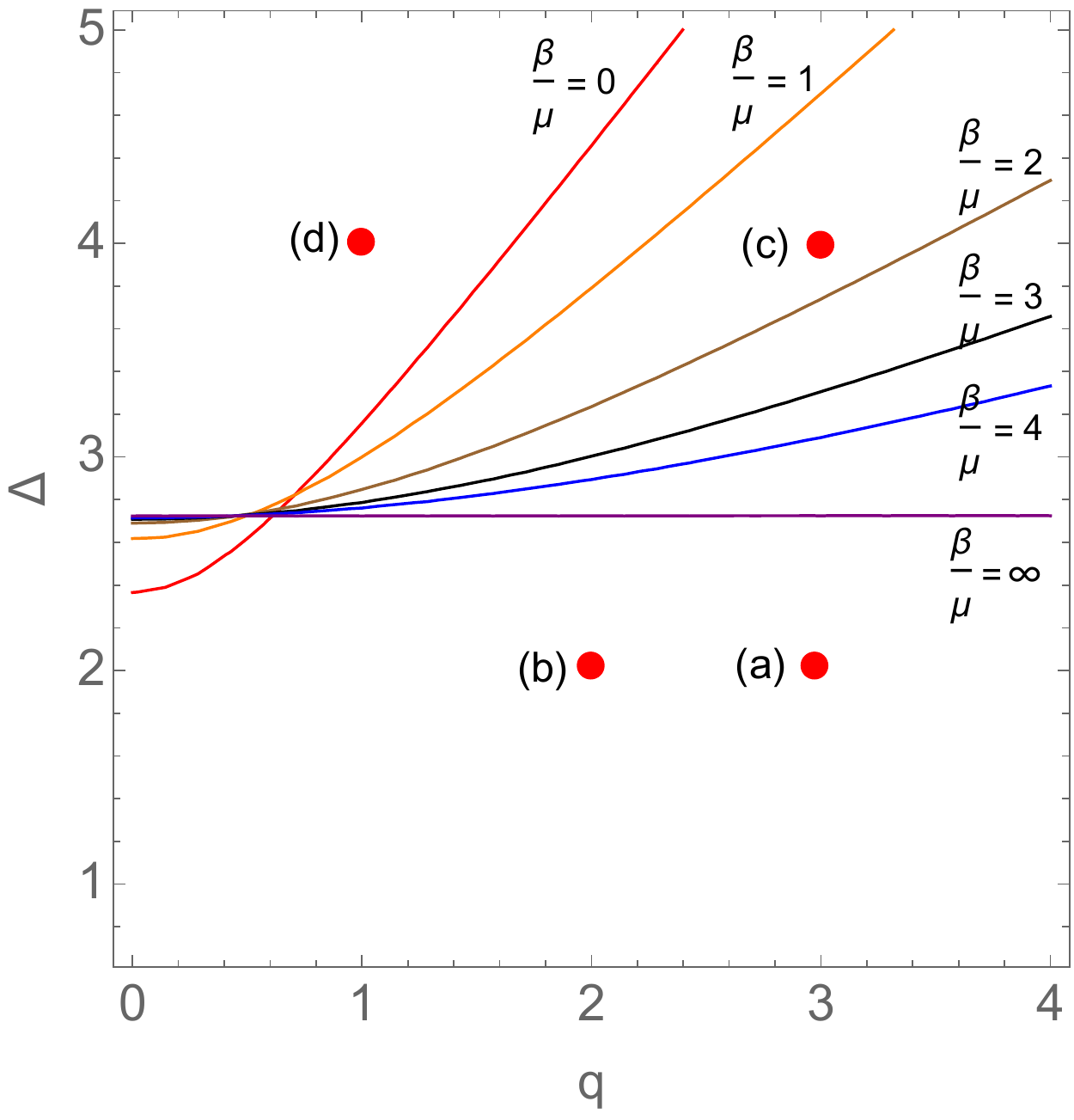} \label{}}
\caption{ Phase boundaries in $q$ and $\Delta$ space. The region above every curves is the normal metal phase. The orange solid curve in (a) and (b) are the same.  }
            \label{Instab}
\end{figure}

Our numerical procedure is as follows.
The equation \eqref{BF1} depends only on four dimensionless quantities: $\Delta$, $q$, $\beta/\mu$, and $T/\mu$. 
For fixed $\beta/\mu$, $\Delta$, and $q$, we shoot from horizon to boundary keeping dialing $T/\mu$. Near boundary, $\phi$ falls off as $r^{\Delta-3}$ and $r^{-\Delta}$. We search for the largest value of $T/\mu$ for which the coefficient of $r^{\Delta-3}$ vanishes. We repeat this procedure for different pairs of values of $\Delta$, and $q$.
Figure \ref{Instab}(a) shows the line of constant critical temperature ($T/\mu$) in $\Delta$-$q$ space when $\beta/\mu = 1$.

In Figure \ref{Instab}(a) there is a special curve(solid one) representing quantum phase transition at $T=0$.  We may understand the quantum phase transition by considering the Breitenlohner-Freedman(BF) bound.
First, the effective mass of scalar field, which is read from \eqref{BF1},  near horizon at zero temperature is  
\begin{equation}
m_{\mathrm{eff}}^2 =\lim_{r \to r_h } \lim_{T \to 0}  \left( m^2    + q^2 g^{tt} A_t^2  \right)= m^2 - \frac{2q^2}{1 + \frac{\beta^2}{\mu^2}} \,. \label{data1}
\end{equation}
Second, the near horizon geometry of an extremal  black brane is given by AdS$_2\times \mathbb{R}^ {2}$ with 
the effective radius of AdS$_2$ given by \eqref{AdS2}
\begin{equation}
L_{2}^2 = \frac{L^2}{6}\left(1+ \frac{\frac{\beta^2}{\mu^2}}{1+ \frac{\beta^2}{\mu^2} }\right) \,.  \label{data2}
\end{equation}
Third, recall that, real scalar field in the AdS$_{d+1}$ space with the radius $L_{d+1}$ is unstable with mass($M$) below the BF bound
\begin{equation}
M^2 L_{d+1}^2 = - \frac{d^2}{4} \,.  \label{data3}
\end{equation}
From these three data \eqref{data1}-\eqref{data3}, we conclude that the scalar field is unstable near horizon if 
\begin{equation} \label{BF2}
\tilde{m}_{\mathrm{eff}}^2 \equiv m_{\mathrm{eff}}^2  L_2^2 =  \left[m^2 - \frac{2q^2}{1 + \frac{\beta^2}{\mu^2}} \right] \left[\frac{L^2}{6}\left(1+ \frac{\frac{\beta^2}{\mu^2}}{1+ \frac{\beta^2}{\mu^2} }\right)\right]  < -\frac{1}{4} \,,
\end{equation}
where $m^2L^2$ must be greater than $-\frac{9}{4}$ for scalar to be stable at boundary AdS$_4$ space.
Alternatively, this result can be obtained from \eqref{BF1}. 
For the scalar field near horizon of the extremal black hole, (\ref{BF1}) becomes  
\begin{align}\label{fluc scalar}
\phi''(\eta) +  \frac{2}{\eta} \phi'(\eta)  -  \frac{\tilde{m}_{\mathrm{eff}}^2}{\eta^2} \phi(\eta)=0~~, 
\end{align}
where $\eta= (r-r_h)$ and $\tilde{m}_{\mathrm{eff}}^2$ is defined in \eqref{BF2}.
The equation (\ref{fluc scalar}) is  an equation for a real scalar field in AdS$_2$ spacetime with the curvature radius of unity.  Thus, we reach the same conclusion as \eqref{BF2} by \eqref{data3}. 

From the inequality \eqref{BF2},  we can infer that qualitatively a larger mass (or $\Delta$)\footnote{The mass $m^2$ is related to the conformal dimension $\Delta$ of an operator in the dual field theory, $m^2 = \frac{\Delta (\Delta - 3)}{L^2}$.   }, smaller $q$, or larger $\beta/\mu$ suppresses instability(superconductivity) at zero temperature.   Quantitative phase boundaries for several $\beta/\mu$ are plotted in Figure \ref{Instab}(b), where the region above (below) the curve is stable or normal phase (unstable or superconducting phase).
It is interesting that one can tune $\beta/\mu$ to trigger a qunatum phase transition. For example, let us consider the system with $q=3$ and $\Delta = 4$, which is (c) in Figure \ref{Instab}(b). When $\beta/\mu$ is small the system is in normal phase, but when $\beta/\mu$ is large the system is in superconducting phase. The transition occurs between $\beta/\mu=1$ and $\beta/\mu=2$.  The system at (d) in Figure \ref{Instab}(b) must be always in the normal phase at zero temperature regardless of $\beta/\mu$. 
However the system with $q=2, \Delta = 2$ and $q=3, \Delta=2$ ((a) and (b) in Figure \ref{Instab}(b) respectively) must be always in the superconducting phase at zero temperature. 
Notice that, in this case, the superconducting transition occurs even at $\mu=0$, i.e. $\beta/\mu = \infty$. It also can be seen in Figure \ref{phase3D}.

\subsection{Superconducting phase}

In the previous subsection we investigated the possibility of superconducting phase at finite temperature and zero temperature. Based on our result on quantum phase transition, Figure \ref{Instab}(b), we may anticipate which values of parameters ($q, \Delta, \beta, \mu$) allow thermal phase transition. In this subsection we want to confirm our anticipation by constructing explicit superconducting background at finite temperature.

\begin{figure}[]
 \centering
   \subfigure[$\Delta =2 , q=3$ ]
   {\includegraphics[width=4.5cm]{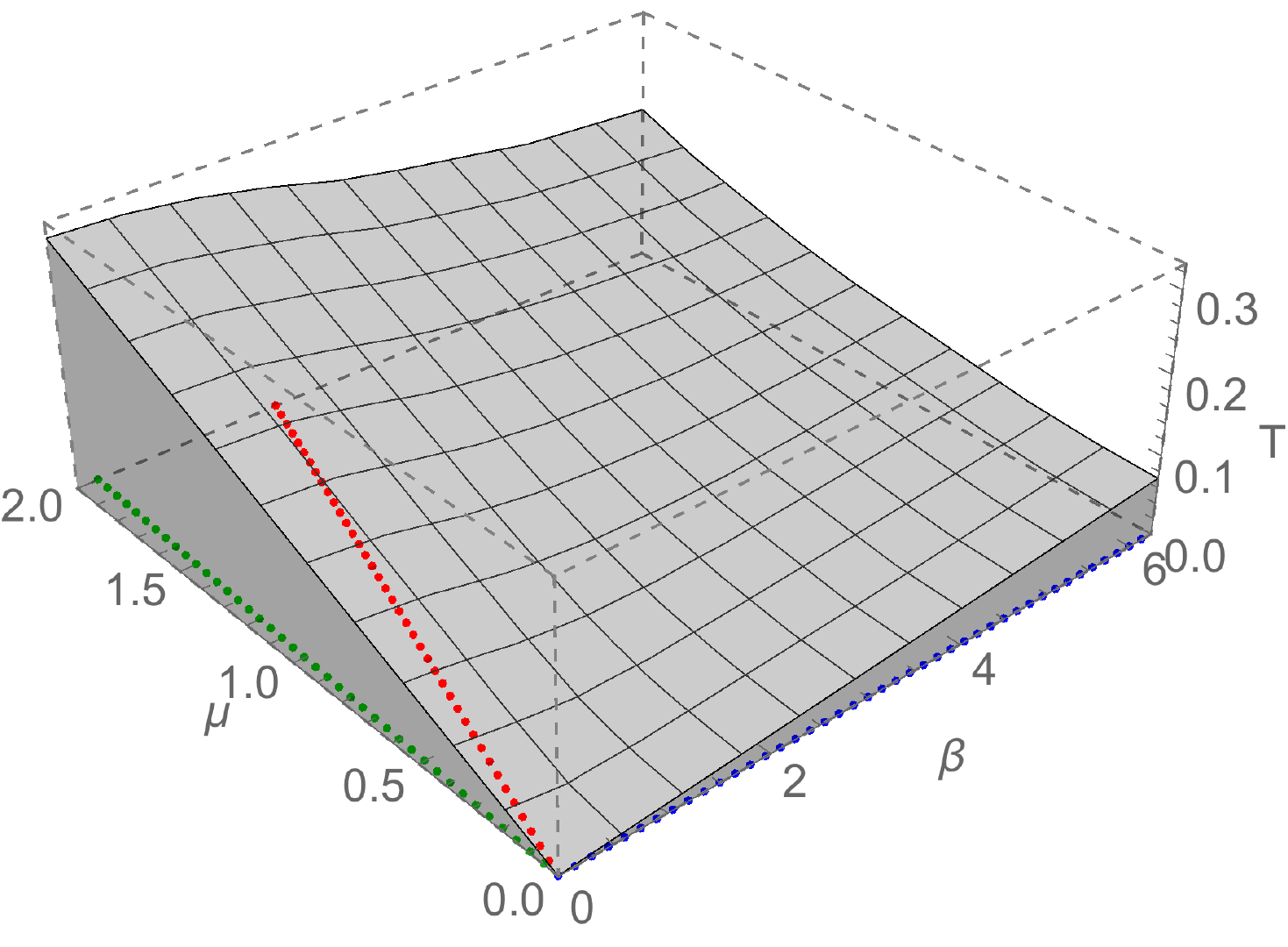} \label{}}  
  \subfigure[$\Delta =2 , q=2$ ]
   {\includegraphics[width=4.5cm]{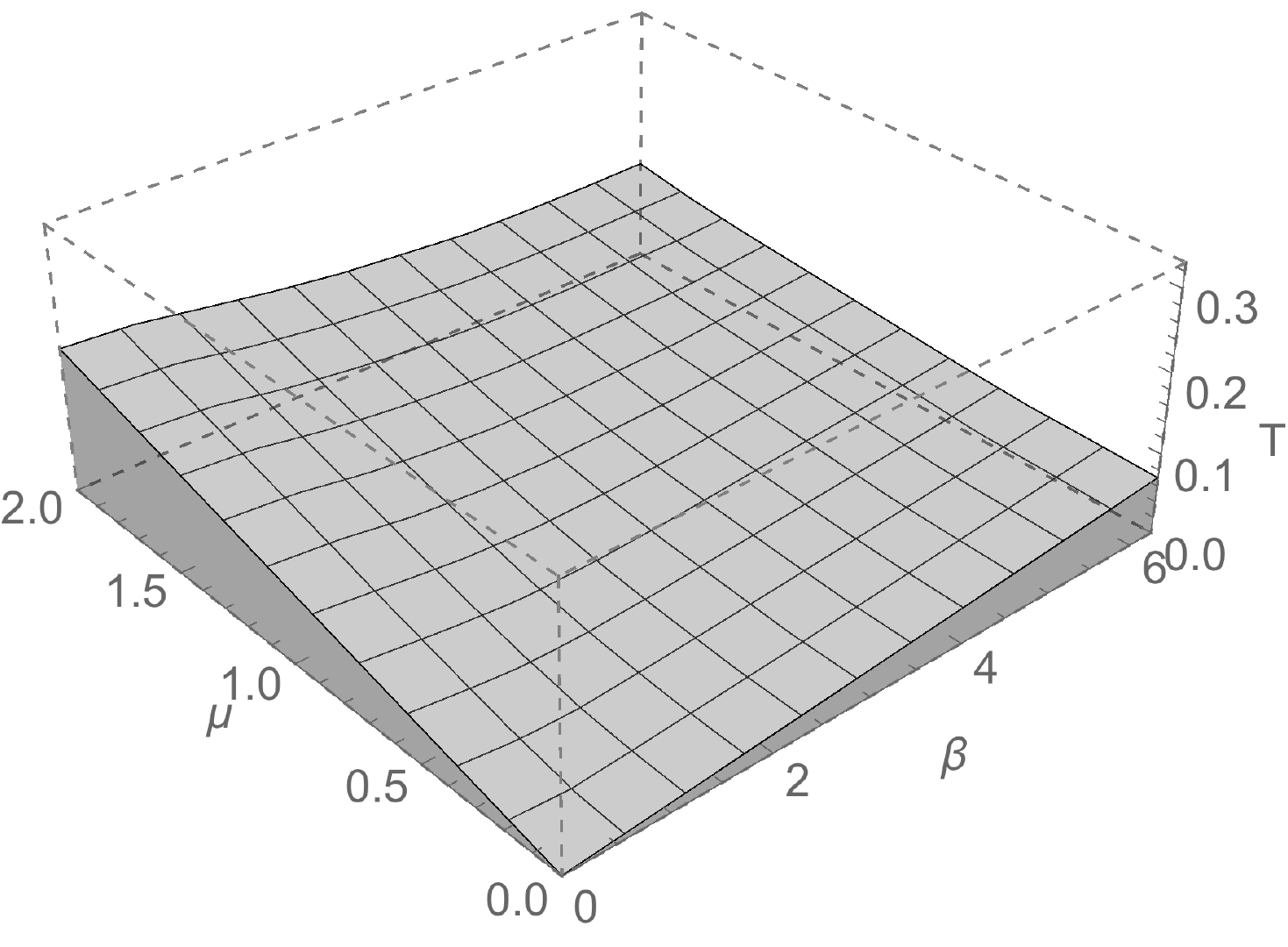} \label{}}   
     \subfigure[$\Delta = 4 , q=3$ ]
   {\includegraphics[width=4.5cm]{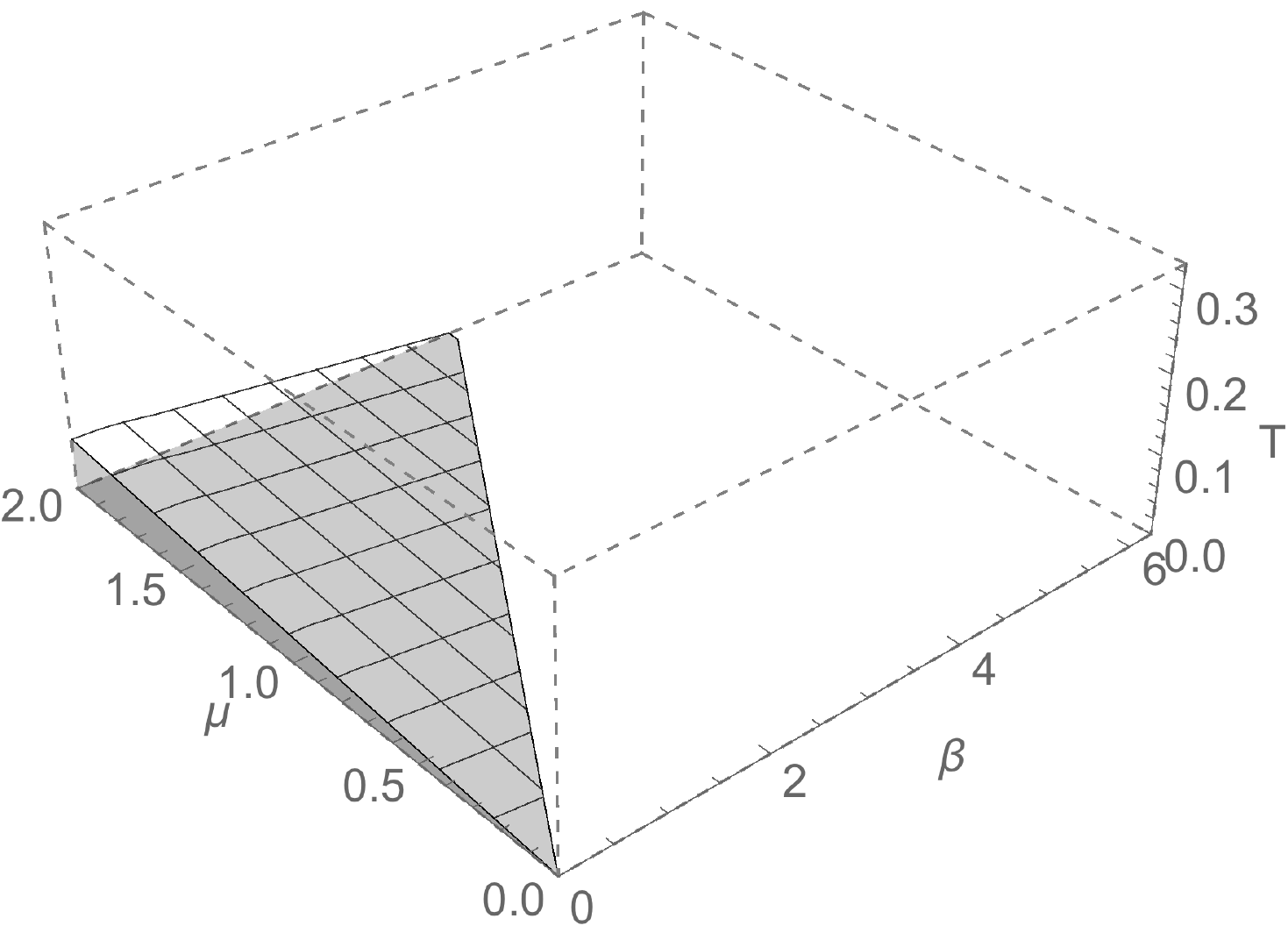} \label{}}
    \subfigure[ plot (a) and (b) together]
   {\includegraphics[width=4.5cm]{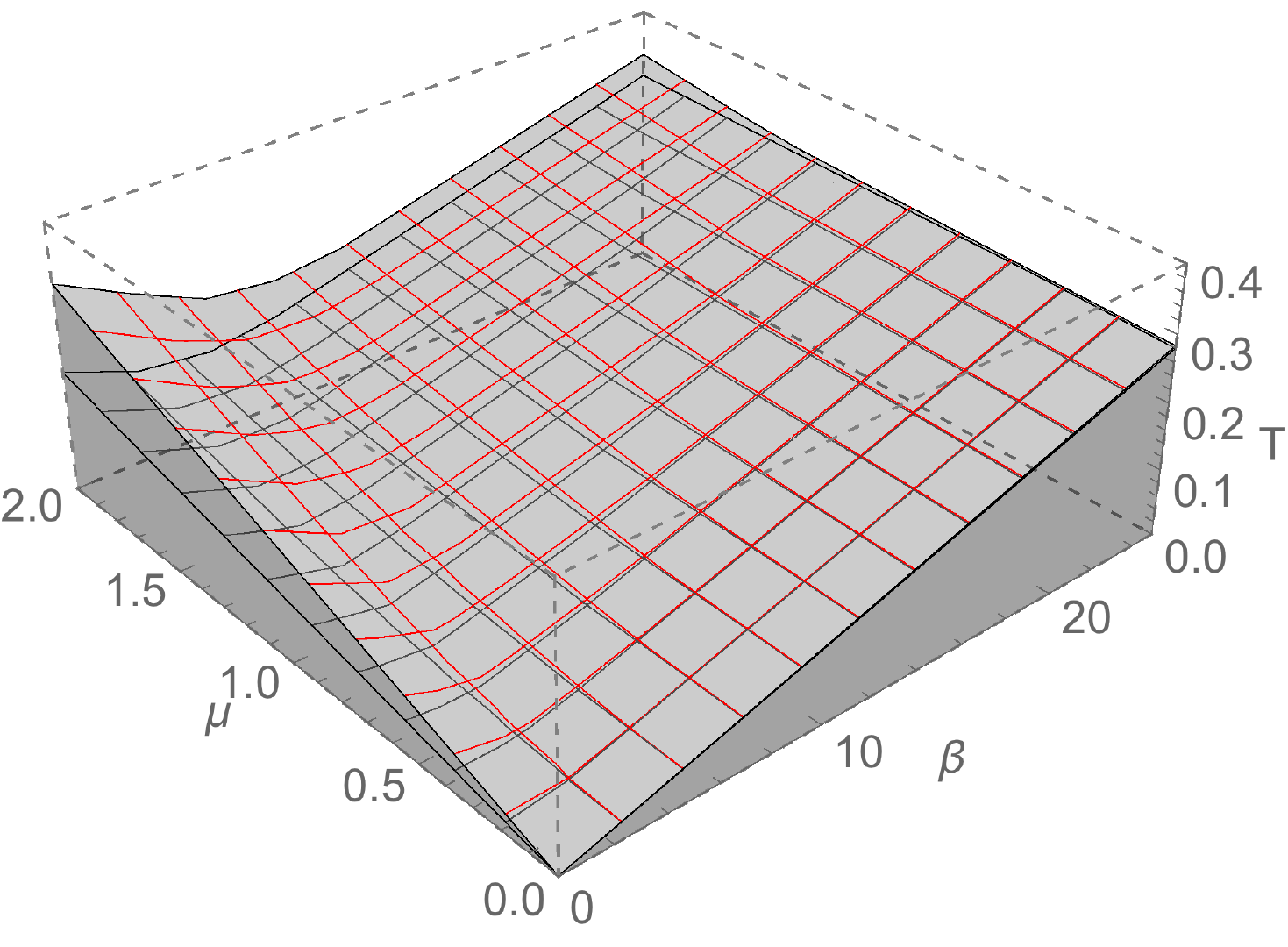} \label{}}  
  \subfigure[plot (c) extended to $-\beta$]
   {\includegraphics[width=4.5cm]{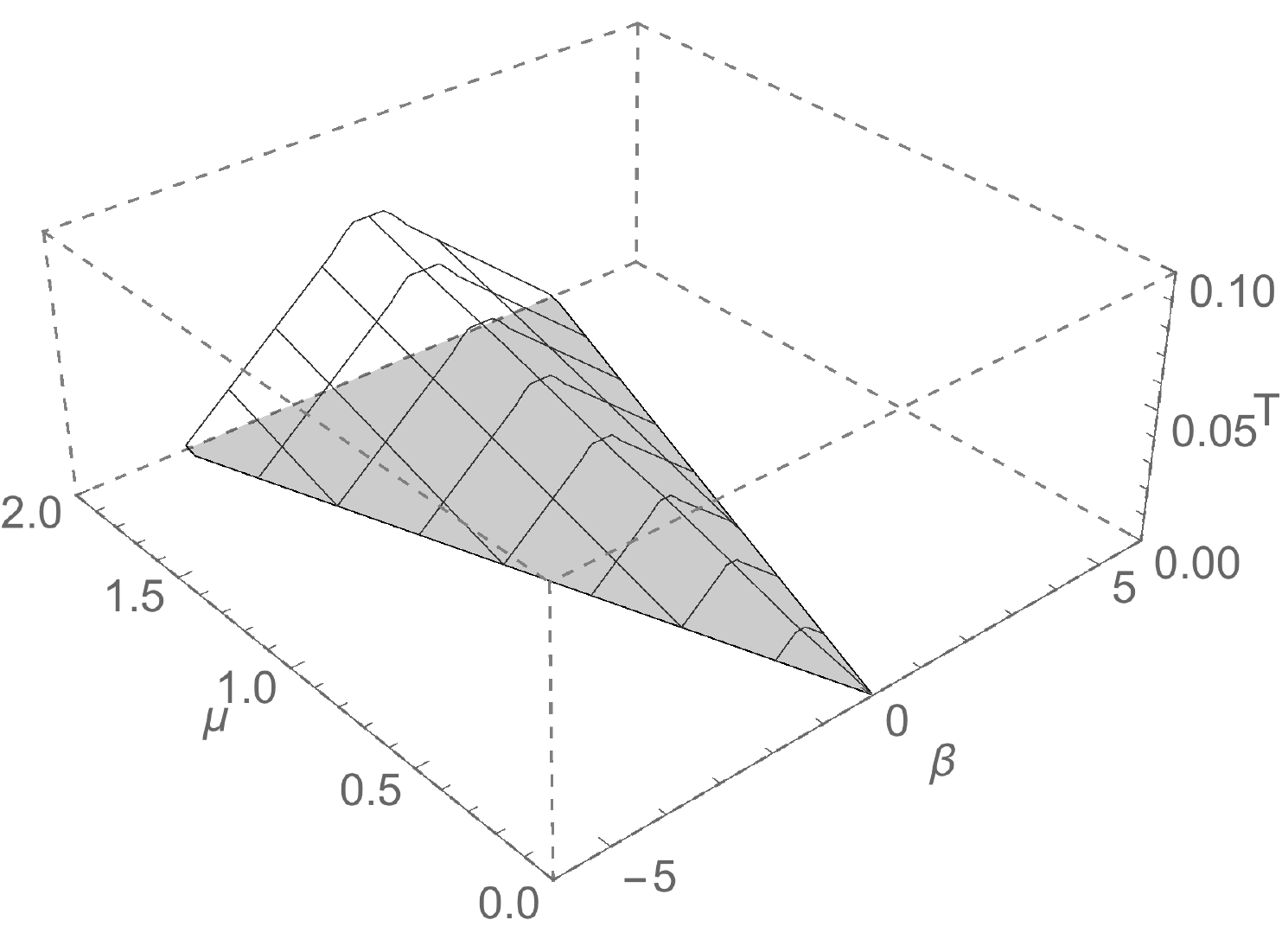} \label{}}   
  \caption{Phase diagrams for point (a),(b), and (c) in Figure \ref{Instab}. The meshed surface is the phase boundary at the critical temperature. 
  Dark region below the surface is superconducting phase while region above the surface is normal phase.}
            \label{phase3D}
\end{figure}

Our numerical analysis is performed as follows. By shooting out from horizon we find a numerical solutions satisfying \eqref{eom5}-\eqref{eom8} and boundary condition $\Phi^{(1)} = 0$ and consider the condensate of the operator of dimension two,  $\langle\mathcal{O}^{(2)}\rangle$. See \eqref{coeff1}. We may choose the different boundary condition, 
$\Phi^{(2)} = 0$, but we will not deal with the case in this paper.  At high temperature we obtain only one solution, which agrees to an analytic solution of normal state  \eqref{normal1}-\eqref{normal4}. At low temperature we find another solution with $\Phi \ne 0$(superconducting phase) in addition to a normal state solution \eqref{normal1}-\eqref{normal4}. In this case  it turns out that the superconducting solution always has a lower grand potential and becomes a ground state. The phase transition  is continuous at a critical temperature($T_c$). Figure \ref{phase3D} shows typical examples of phase diagrams for three points (a),(b), and (c) in Figure \ref{Instab}(b).  The three dimensional information in Figure \ref{phase3D} may be summarized in a two dimensional plot, for example, in the plane of dimensionless quantities $T/\beta$ and $\mu/\beta$. In practice, we have obtained such two dimensional plots first and rescaled them to make three dimensional plots, where $\mu, \beta,$ and $T$ have the same unit of energy. Three dimensional plots would be more convenient to represent overall features, even though all information can be compressed in two dimensional plots.

Let us start with the point (a) and (b) in Figure \ref{Instab}(b). They are always in superconducting phase  for {\it{all}} $\beta$ and $\mu$ at zero temperature. As temperature increases we expect that the system undergoes a phase transition from superconducting phase to normal phase. Our numerical analysis confirms it and the phase diagram is  shown in Figure \ref{phase3D}(a)(b), where the meshed surface is the phase boundary at the critical temperature. Dark region below the surface is superconducting phase while region above the surface is normal phase.   
Figure \ref{phase3D}(a)(b) focuses on the phase structure for small $\beta$.  In Figure \ref{phase3D}(d) we extend $\beta$ axis of Figure \ref{phase3D}(a)(b)  to larger values and combine them for comparison,  where Figure \ref{phase3D}(a) is red.  The red mesh is above the black mesh, which means that a large $q$ enhances 
superconductivity, as at the zero temperature in Figure \ref{Instab}.
However, the phase transition line coincides at $\mu=0$, because the effect of $q$ enters in the combination of $q \mu $ as shown in  \eqref{BF1}. 

Notice that the superconducting phase exists even when $\mu=0$. In the HHH model~\cite{Hartnoll:2008kx} the phase transition is understood as a competition between $\mu$ and $T$ and in this case it is a competition between $\beta$ and $T$. The `pairing mechanism' of two cases must be different, because when $\beta=0$ it would be due to a particle-particle pair, while when $\mu=0$ it would be due to a particle-anti-particle pair also interacting with $\beta$, which may be interpreted as `impurity'~\cite{Kim:2014bza}. In general, at finite $\mu$ and $\beta$, two mechanisms will compete. 

This competition is reflected on the phase boundary surfaces in Figure \ref{phase3D}.  
In words, the dependence of the critical temperature on $\beta/\mu$ is not monotonic. the critical temperature decreases when $\beta/\mu$ is small and increases when $\beta/\mu$ is large. In graphics, see the line at $\mu=2$ in Figure \ref{phase3D}(d) or the lines at $\mu=0$ and $\mu=2$ in Figure  \ref{phase3D}(a). 
It is different from the previous studies. In Q-lattice model~\cite{Ling:2014laa,Andrade:2014xca} and single scalar model~\cite{Koga:2014hwa}  the critical temperature decreases as momentum relaxation effect increases while in ionic lattice model~\cite{Horowitz:2013jaa} the critical temperature increases  monotonically as lattice effect increase.

In Figure \ref{Instab}(b), the point (c) is different from (a) and (b), in that the system at (c) at zero $T$ could be in superconducting phase or normal phase depending on the value of $\beta/\mu$; the critical value is around $\beta/\mu \approx 1.5$.  It is confirmed by considering the finite temperature transition in Figure \ref{phase3D}(c), where the transition close to $T=0$ occurs around at the line of $\beta/\mu \approx 1.5$\footnote{We have not yet obtained the
solution at zero $T$. The data for the plot is numerically computed up to $T/\mu \sim 0.01$.  
It will be interesting to construct and analyse the zero $T$ limit solutions more precisely, for example, following \cite{Horowitz:2009ij}.}.
If we include negative values of $\beta$, we obtain the phase diagram Figure \ref{phase3D}(e).  It looks similar to the superconducting dome in cuprate superconductor phase diagram when we identify $\beta$ with a doping parameter.
Note that $\beta$ is a tunable continuous free parameter  in the solution, while $m$ and $\Delta$ is fixed in the action. 
We also see large $\Delta$ suppresses superconductivity by comparing Figure \ref{phase3D}(a) and (c).

 \begin{figure}[]
 \centering
   {\includegraphics[width=9cm]{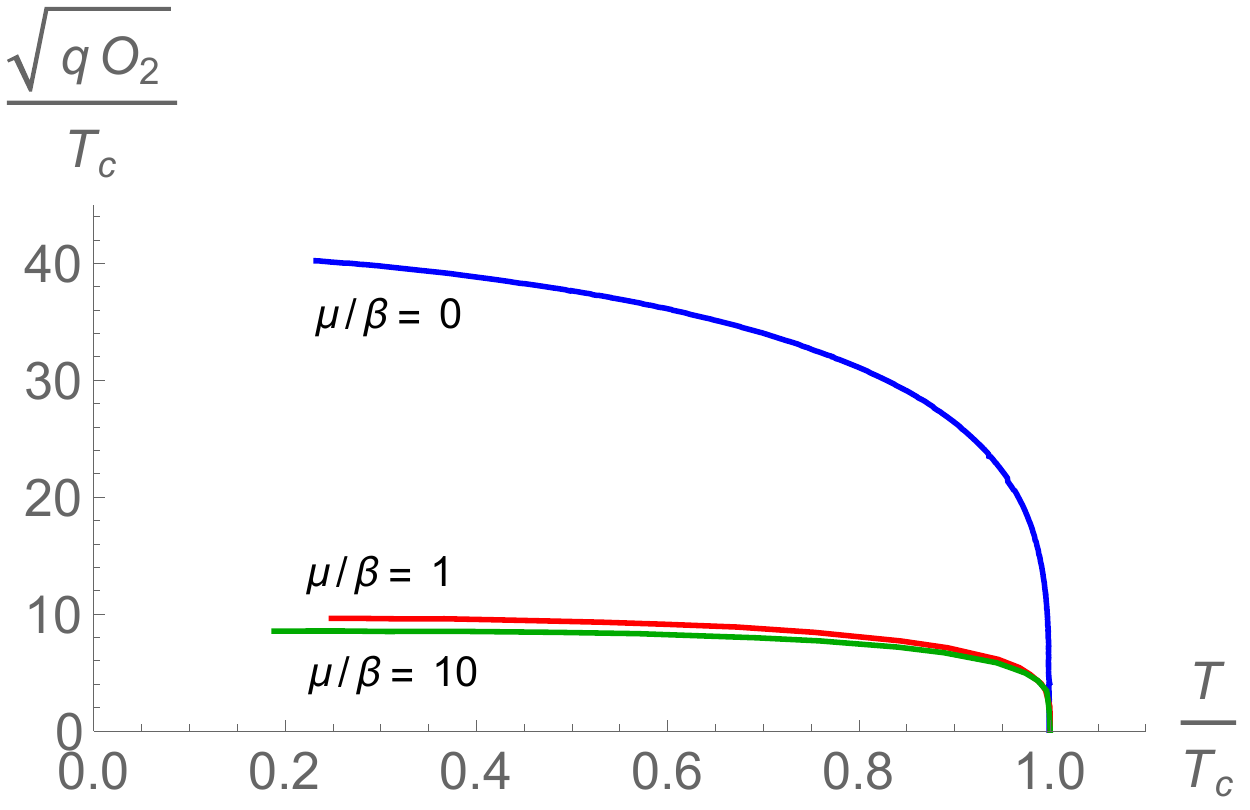} \label{}}
  \caption{ Condensate for three values of $\mu/\beta$ and $\Delta =2 , q=3$, which is the case of Figure \ref{phase3D}(a). The color here matches the color of the lines in Figure \ref{phase3D}(a). In other words, we compute condensate along the vertical line(temperature) standing on the colored-lines in Figure \ref{phase3D}(a)  }
            \label{Condensate}
\end{figure}

In superconducting phase there is finite condensate, $\Phi^{(2)}$, which is an order parameter. 
For example, in Figure \ref{Condensate}, we show the condensate as a function of temperature.  At the critical temperature condensate starts forming and increases continuously as temperature goes down.  At very small temperature our numerics becomes not reliable and we did not plot in that range.  The plot is for $\Delta =2 , q=3$ which is the case of Figure \ref{phase3D}(a). Three values of $\mu/\beta$ are chosen: $\mu/\beta=0$(blue), $\mu/\beta=1$(red), and $\mu/\beta=10$(green).  
The color here matches the color of the lines in Figure \ref{phase3D}(a). In other words, we compute condensate along the vertical line(temperature) standing any point on the colored-lines in Figure \ref{phase3D}(a).
The condensate increases as $\beta$ increases, which agrees to an anisotropic case \cite{Koga:2014hwa}. When $\beta \rightarrow \infty$ the condensate approaches to the finite upper bound, 
while when $\beta \rightarrow 0$  it approaches to the lower bound.

We finish this subsection by discussing on the on-shell action of the ground state.
To calculate a thermodynamic potential for the black hole solutions we calculate the on-shell Euclidean action($S^E$) by  analytically continuing to Euclidean time($\tau$)
\begin{equation}
t= -i \tau\,, \qquad S^{\mathrm{E}} = -i S_{\mathrm{ren}} \,,
\end{equation}
where $S_{\mathrm{ren}}$ consists of four-dimensional  $S_{\mathrm{bulk}}$ and three dimensional $S_{\mathrm{bdy}}$:  
\begin{equation} \label{renact}
S_{\mathrm{ren}} \equiv \underbrace{S_{\mathrm{HHH}} + S_\psi}_{\equiv S_{\mathrm{bulk}}}  + \underbrace{S_{\mathrm{GH}} + S_{\mathrm{ct}}}_{\equiv S_{\mathrm{bdy}}}  \,.
\end{equation}
The first three terms are defined in \eqref{HHH}, \eqref{psi} and \eqref{GH} respectively, and the last term is the counter term for holographic renormalisation \cite{Hartnoll:2008kx}:
\begin{equation}
S_{\mathrm{ct}} =  \int \dd^3 x \sqrt{-\gamma}  \left( - \frac{4}{L}    +   \frac{L}{2} \nabla \psi_I \cdot \nabla   \psi_I    +  \left( \eta_1  (  \Phi^*  n^M  \partial_M \Phi +\Phi  n^M  \partial_M \Phi^*   )     +\eta_2  |\Phi|^2/L  \right)     \right) \,,
\end{equation}
which cancels the divergence of the bulk action.  To fix $\Phi^{(1)}$ on the boundary  we choose $(\eta_1, \eta_2) =(0,-1)$ while to fix $\Phi^{(2)}$ we choose $(\eta_1, \eta_2) =(1,1)$. $n^M =(0,0,0,  \sqrt{\mathcal G (r)} )$ is an outgoing normal vector.  See appendix \ref{apa} for more details. 

Let us first consider the Euclidean bulk action
\begin{align}
S_{\mathrm{bulk}}^\mathrm{E} = - \int \dd^4 x \sqrt{-g} ~\mathcal L_{\mathrm{bulk}} \,, 
\end{align}
which defines $\mathcal L_{\mathrm{bulk}}$.
It can be computed following \cite{Hartnoll:2008kx}.
The $xx$-component of the Einstein equation gives a useful relation:
 \begin{align}
G_{xx}  = \frac{1}{2} r^2(\mathcal L_{\mathrm{bulk}} - R)   + \frac{1}{2} \beta^2~,
\end{align}
where $G_{\mu\nu}$ is the Einstein tensor.  The trace of the Einstein equation yields 
\begin{align}
G^t_t + G^r_r  +  \frac{\beta^2}{r^2}  -  R  +  \mathcal L_{\mathrm{bulk}} = -R \,.
\end{align}
Thus the bulk Lagrangian is
\begin{align}
\mathcal L_{\mathrm{bulk}} = - G^t_t -G^r_r - \frac{\beta^2}{r^2}   \,.
\end{align}
In superconducting phase
\begin{align}
S_{\mathrm{bulk}}^E=- \int \dd^3 x \int_{r_h}^{\infty} dr \left\{   \left(  -2 r g(r)  e^{-\chi/2}  \right)'    -\beta^2  e^{-\chi/2}    \right\}~~.
\end{align}
After adding $S^E_\mathrm{bdy}$ the total on-shell action is finite and reads
\begin{equation}
S^{E} =   S_{\mathrm{bulk}}^E +S^E_\mathrm{bdy} =  \int_0^{  1/T  } \dd \tau  \int \dd x_1 \dd x_2  \mathcal{W}  = \frac{V_2}{T} \mathcal W~~,
\end{equation}
where we consider a homogeneous and equilibrium system. Thus $V_2$ is the volume of the system and the length of  the Euclidean time circle is identified with inverse temperature($T$).  $\mathcal W$ is a thermodynamic potential per unit volume:
\begin{equation}
\mathcal W = \tg^{(1)} -\beta^2 r_h   -  \beta^2 \int_{r_h}^\infty \dd r  \left(e^{-\chi/2} - 1\right)  -\frac{ 2 \left(2-3  \eta_1+ \eta_2 \right) \Phi ^{(1)} \Phi^{(2)} }{ L^2}  \,.
\end{equation}
In our case, one of $\Phi^{(1)}$ or $\Phi^{(2)}$ is zero and $\mathcal W$ is simplified 
\begin{equation}
\mathcal W = \tg^{(1)} -\beta^2 r_h   -  \beta^2 \int_{r_h}^\infty \dd r  \left(e^{-\chi/2} - 1\right) \,. \label{pgrand}
\end{equation}

In normal sate, where $\tg^{(1)} = -m_0$ and $\chi=0$, \eqref{pgrand} becomes
\begin{equation} \label{WW}
\mathcal W = \frac{r_h}{4 } \left(- 4 r_h^2-\mu ^2  
-2 \beta ^2\right) \,.
\end{equation}
With the relations for energy density($\epsilon$), charge density ($\rho$), and entropy density($s$)
\begin{equation} \label{rel1}
\epsilon = \left< T_{tt} \right> = 2 m_0 \,,  \quad \rho = \left< J^t \right> = \mu r_h \,, \quad s= 4\pi r_h^2 \,,
\end{equation}
derived from (\ref{boundary tensors})  and the definition of the Bekenstein-Hawking entropy,
the equation \eqref{WW} becomes the expression for the grand potential($\Omega$)
\begin{equation}
\mathcal W = \epsilon - T s -\mu \rho = \Omega \,, \label{Wdensity}
\end{equation}
which agrees to {\cite{Andrade:2013gsa}.
 
In superconducting phase, 
\begin{equation} \label{rel1}
\epsilon = \left< T_{tt} \right> = - 2 \tg^{(1)} \,, \quad  \rho = \left< J^t \right> = A^{(1)} \,, \quad s= 4\pi r_h^2 \,, 
\end{equation}
where $\tg^{(1)}$ and $A^{(1)}$ are numerical  values.
Lack of the analytic relation such as  \eqref{rel1} make it difficult to check if $\mathcal{W} = \Omega$.
Therefore, we numerically checked and found that $\Omega=  \epsilon - T s -\mu \rho = \tg^{(1)} -\beta^2 r_h \ne \mathcal{W}$.
However, as far as the phase diagram is concerned, this difference does not matter.  
It turns out $\mathcal{W} < \Omega$  in superconducting phase so we may use $\Omega$ as our criteria for phase transition. To study thermodynamical quantities it will be important to understand physics of the difference between $\mathcal{W}$ and $\Omega$. We leave it for future study. 

\section{Optical conductivity} \label{sec3}

In this section we study electric($\sigma$), thermoelectric($\alpha$), thermal($\bar{\kappa}$)  conductivity by considering small fluctuations of relevant gauge, metric, scalar fields around the normal and superconducting background we obtained in the previous section. From here on, we set $L=1$ and use the scaled variables \eqref{scaling3} without tilde.

\subsection{Fluctuations for optical conductivity: equations and on-shell action}

Electric conductivity is related to a small bulk gauge field fluctuation $\delta A_x(t,r)$
\begin{equation} \label{flucA}
 \delta A_x(t,r) = \int^{\infty}_{-\infty} \frac{\dd \omega}{2\pi}  e^{-i\omega t}  a_{x}(\omega,r) \,,
\end{equation}
of which boundary dual operator is electric current.  The fluctuation is chosen to be independent of $x$ and $y$, which is allowed since all the background fields affecting the equations of motion are independent of $x$ and $y$. Because of rotational symmetry in $x$-$y$ plane, it is enough to consider $ \delta A_x$.
The gauge field fluctuation($\delta A_x(t,r)$) sources to metric($\delta  g_{tx}(t,r) $) and scalar field($\delta \psi_1(t,r)$) fluctuation
\begin{align}
\delta g_{tx}(t,r) &=  \int^{\infty}_{-\infty} \frac{\dd \omega}{2\pi} e^{-i\omega t} r^2 h_{tx}(\omega,r)\,,  \label{flucG} \\ 
\delta \psi_1(t,r) &= \int^{\infty}_{-\infty} \frac{\dd \omega}{2\pi} e^{-i\omega t}  \xi (\omega,r) \,, \label{flucPsi}
\end{align}
and all the other fluctuations can be decoupled. Notice that $h_{tx}(\omega, r)$ is defined to approach constant as $r$ goes to infinity.

In momentum space, the linearized equations of motion around the background \eqref{ansatz1} are derived from \eqref{eom1}-\eqref{eom4}:
\begin{align}
&a_x''+ \left(\frac{\tg'}{\tg}-\frac{\chi'}{2}\right)a_x' + \left( \frac{\omega ^2}{\tg^2} e^{\chi}- \frac{2 q^2 \Phi^2}{\tg} \right) a_x +\frac{r^2 e^{\chi } A_t' }{\tg}h_{tx}'=0 \,,  \label{feom1} \\
&h_{tx}' +\frac{ A_t'}{r^2} a_x +\frac{i \beta  \tg e^{-\chi }}{r^2 \omega } \xi'=0 \,, \label{feom2} \\
&\xi ''+\left(\frac{\tg'}{\tg}-\frac{\chi'}{2}+\frac{2}{r}\right) \xi'  -\frac{i \beta  \omega  e^{\chi } }{\tg^2} h_{tx} +\frac{\omega ^2  e^{\chi }}{\tg^2} \xi  =0 \,, \label{feom3}
\end{align}
where $\tg, \chi, A_t, \Phi$ are background field obtained in the previous section.  For normal phase we have the analytic solutions (\eqref{normal1}-\eqref{normal4}) but for superconducting phase we have it numerically.

We solve these equations with two boundary conditions: incoming boundary conditions at the black hole horizon and the Dirichlet boundary conditions at the boundary. 
First, near the black hole horizon ($r \rightarrow 1$) the solutions are expanded as
\begin{align}
h_{tx} &= (r-1)^{\nu_\pm  +  1} \left(  h_{tx}^{(I)}   + h_{tx}^{(II)}(r-1) + \cdots    \right) \,, \label{near1} \\
a_{x} &= (r-1)^{\nu_\pm  } \left(   a_{x}^{(I)}   + a_{x}^{(II)}(r-1) + \cdots    \right)\,, \\
\xi &= (r-1)^{\nu_\pm  } \left(  \xi^{(I)}   + \xi^{(II)} (r-1) + \cdots    \right)\,, \label{near2}
\end{align} 
where $\nu_\pm = \pm i\omega \frac{e^{    \chi(1)/2   } }{\mathcal{-G}'(1)}$
and the incoming boundary condition corresponds to $\nu = \nu_+$.   
Next, near the boundary ($r \rightarrow \infty$) the asymptotic solutions read
\begin{align} \label{nearb}
h_{tx} &=   h^{(0)}_{tx} + \frac{1}{r^2} h^{(2)}_{tx} + \frac{1}{r^3}h_{tx}^{(3)}+\cdots \,,  \\
a_x&=a_x^{(0)} + \frac{1}{r}a_x^{(1)}+ \cdots \,, \qquad  \quad \\
 \xi &= \xi^{(0)} + \frac{1}{r^2} \xi^{(2)}+ \frac{1}{r^3}\xi^{(3)} + \cdots \,,
\end{align}
and we fix the values of the leading terms as boundary conditions.

Plugging the solutions into the renormalized action \eqref{renact}, we have a quadratic order on-shell action 
\begin{equation}
\begin{split}
S^{(2)}=\int d^3 x  & \left[  h_{tx} \left(-\frac{1}{2} r^2 e^{\chi /2} \mathit{a}_x A_t'+\frac{\beta  \dot{\xi } r^2 e^{\chi /2}}{2 \sqrt{\mathcal{ G }}}+\frac{1}{2} r^4 e^{\chi /2} h_{tx}'\right) \right. \\ 
&+e^{\chi /2} h_{tx}^2 \left(  2 r^2 - \frac{2r^4}{\sqrt{\mathcal{ G }}}+\frac{r^4  \Phi ^2 }{2 \sqrt{\mathcal{ G }}}\right) \\
&\left. -\frac{\beta  \xi  r^2 e^{\chi /2} \dot{h}_{tx}}{2 \sqrt{\mathcal{ G }}}-\frac{1}{2} \mathcal{ G }  e^{-\frac{\chi }{2}} \mathit{a}_x \mathit{a}_x'-\frac{1}{2} \mathcal{ G } \xi  r^2 e^{-\frac{\chi }{2}} \xi '+\frac{\xi  r^2 e^{\chi /2} \ddot{\xi }}{2 \sqrt{\mathcal{ G }}}   \right] \,,
\end{split}
\end{equation} 
where we discarded the contribution from the horizon as the prescription for the retarded Green's function \cite{Son:2002sd}. 
In particular, with the spatially homogeneous ansatz \eqref{flucA}-\eqref{flucPsi}, the quadratic action in momentum space yields 
\begin{equation} \label{onshell2}
S^{(2)}_{\mathrm{ren}} = \frac{V_2 }{2} \int_0^\infty \frac{\dd \omega}{2\pi}    \left(   -  \rho  a_x^{(0)} h_{tx}^{(0)}  +   2\mathcal{G}^{(1)}  h_{tx}^{(0)}  h_{tx}^{(0)}  + a_x^{(0)} a_x^{(1)} -3 h_{tx}^{(0)} h_{tx}^{(3)}    + 3 {\xi}^{(0)} \xi^{(3)}   \right) \,,
\end{equation}
where $V_2$ is the two dimensional spatial volume $\int \dd x \dd y$ and we omit the term proportional to $ \Phi ^{(1)} \Phi ^{(2)} h_{tx}^{(0)}h_{tx}^{(0)}$ since we are studying the case with  $\Phi ^{(1)} =0 $. The range of $\omega$ is chosen to be positive following the prescription in  \cite{Son:2002sd}.

The on shell action \eqref{onshell2} plays a role of the generating functional for two-point Green's functions sourced by $a_x^{(0)}, h_{tx}^{(0)}$, and ${\xi}^{(0)}$. We may simply read off part of the two point functions from the first two terms in \eqref{onshell2}.  The other three terms are nontrivial and we need to know the dependence of $\{a_x^{(1)}, h_{tx}^{(3)}, \xi^{(3)}\}$ on $\{a_x^{(0)}, h_{tx}^{(0)},{\xi}^{(0)}\}$.  However, thanks to linearity of equations \eqref{feom1}-\eqref{feom3}, we can always find out the linear relation between $\{a_x^{(1)}, h_{tx}^{(3)}, \xi^{(3)}\}$ and $\{a_x^{(0)}, h_{tx}^{(0)},{\xi}^{(0)}\}$. 
We will first explain our numerical method to find such a relationship in a more general setup in the following subsection and continue the computation in that setup. 

\subsection{Numerical method}

A systematic numerical method for a system with multi fields and constraints were developed in \cite{Kim:2014bza} based on  \cite{Amado:2009ts,Kaminski:2009dh}. We summarise it briefly and refer to \cite{Kim:2014bza} for more details. 

To develop a systematic method in a general setup, let us start with $N$ fields  $\Phi^a(x,r)$, ($a=1,2,\cdots, N$), which satisfy a set of coupled $N$ independent second order diffrential equations:
\begin{equation}
\Phi^a(x,r) = \int \frac{\dd^d k}{(2\pi)^d}  e^{-ikx}  r^p \Phi^a_k(r)\,, \label{newphi}
\end{equation}
where the index $a$ may include components of higher spin fields. For convenience, $r^p$ is multiplied such that the solution  $\Phi^a_k(r) $ goes to constant at boundary. For example, $p=2$ in \eqref{flucG}.

Near horizon($r=1$), solutions can be expanded as
\begin{equation} \label{incoming}
\Phi^a(r) = (r-1)^{\nu_{a\pm}} \left( \varphi^{a} + \tilde{\varphi}^{a} (r-1) + \cdots \right) \,,
\end{equation}
where we omitted the subscript $k$ for simplicity and $\nu_{a\pm}$ correspond to incoming/outgoing boundary conditions. To compute the retarded Green's function we choose the incoming boundary condition \cite{Son:2002sd},  fixing $N$ initial  conditions. The other $N$ initial conditions, denoted by $\varphi^{a}_{i}$, ($i=1,2,\cdots, N$),  can be chosen, for example, as
\begin{equation} \label{init}
\begin{pmatrix}
    \varphi^{a}_{1} \ & \varphi^{a}_{2}\ & \varphi^{a}_{3}\ &  \ldots \ &  \varphi^{a}_{N}
\end{pmatrix}
   =
\begin{pmatrix}
    1 & 1& 1&  \ldots & 1 \\
    1 & -1& 1 & \ldots & 1 \\
   1 & 1& -1 & \ldots & 1 \\
    \vdots & \vdots & \vdots  & \ddots & \vdots \\
    1 & 1 & 1 & \ldots & -1
\end{pmatrix} \,.
\end{equation}
 Every  column vector $\varphi_i^a$   yields a   solution, denoted by ${\Phi}_i^a(r)$, which is expanded as
\begin{equation}
\Phi_i^a(r)  \rightarrow   \mathbb{S}_{i}^{a}  + \cdots +  \frac{\mathbb{O}_{i}^{a}}{r^{\delta_a}}  + \cdots   \qquad (\mathrm{near\ boundary})\,,
\end{equation}
where  $\mathbb{S}_i^a$ are the {\it{sources}}(leading terms) of $i$-th  solution  and  $\mathbb{O}_{i}^{a}$ are the {\it{operator }}expectation values corresponding to sources($\delta_a \ge 1$).  Notice that $\mathbb{S}$ and $\mathbb{O}$ can be written as regular matrices of order $N$, where the superscript $a$ runs for row index and the subscript $i$ runs for column index.

Since all $N$ solutions $\{\Phi_{i}^{a}\}$ becomes a basis set, a general solution yields
\begin{align} \label{GS} 
\Phi^a(r) = \Phi_{i}^{a}(r) c^i &\rightarrow   \mathbb{S}_{i}^{a} c^i  + \cdots +  \frac{\mathbb{O}_{i}^{a}c^i }{r^{\delta_a}}  + \cdots
 \qquad (\mathrm{near\ boundary})   \\
 &\equiv J^a + \cdots + \frac{R^a}{r^{\delta_a}} + \cdots \,,
\end{align}
with  real constants $c^i$'s. For any given $J^a$ we can always find $c^i$  \,  
\begin{equation}
 c^i  = (\mathbb{S}^{-1})^{i}_{a} J^a \,,
\end{equation}
and the corresponding response $R^a$ is expressed as
\begin{equation} \label{response1}
R^a =  \mathbb{O}_{i}^{a} c^i =  \mathbb{O}_{i}^{a} (\mathbb{S}^{-1})^{i}_{b} J^b \,.
\end{equation} 

A general on-shell quadratic action in momentum space has the form of
\begin{equation} \label{sb}
S_{\mathrm{ren}}^{(2)}  
= \frac{1}{2} \int \frac{\dd^d k}{(2\pi)^d}  \left[ J_{-k}^a \mathbb{A}_{a b}(k) J_k^b
+  J_{-k}^a \mathbb{B}_{a b}(k) {R_k^b} \right]  ,
\end{equation}
where $\mathbb{A}$ and $\mathbb{B}$ are regular matrices of order $N$. 
In matrix notation, $J^a_{-k}$ can be understood as a row matrix. 
For example, the action \eqref{onshell2} can be written in the form \eqref{sb} with
\begin{align}
J^a =
\begin{pmatrix}
    a_x^{(0)}  \\
    h_{tx}^{(0)} \\
   \xi^{(0)} \\
\end{pmatrix}\,, \quad
R^a =
\begin{pmatrix}
    a_x^{(1)}  \\
    h_{tx}^{(3)} \\
   \xi^{(3)} \\
\end{pmatrix}\,, \quad
\mathbb{A} = \begin{pmatrix}
    0 & \ -\rho \ & 0  \\
    0 & \ 2\tg^{(1)} \ & 0  \\
   0 & 0 & 0  \\
\end{pmatrix}\,, \quad
   \mathbb{B}= \begin{pmatrix}
 1 & 0 & 0 \\
 0 & -3 & 0 \\
 0 & 0 & 3 \\
\end{pmatrix} \,,
\end{align}
where the index $\omega$ is suppressed. 
With \eqref{response1}  the action  \eqref{sb}   becomes
\begin{equation} \label{Gab}
\begin{split}
S_{\mathrm{ren}}^{(2)} 
 &= \frac{1}{2}  \int_{\omega \ge 0} \frac{\dd^d k}{(2\pi)^d}  \left[ J_{-k}^a \left[\mathbb{A}_{a b}(k) + \mathbb{B}_{ac}\mathbb{O}_{i}^{c} (\mathbb{S}^{-1})^{i}_{b}(k)\right]J_k^b  \right] \\
& \equiv \frac{1}{2}  \int_{\omega \ge 0} \frac{\dd^d k}{(2\pi)^d}  \left[ J_{-k}^a  G_{ab}^R J_k^b  \right] \,,
\end{split}
\end{equation}
where the range of $\omega$ is chosen to be positive following the prescription in  \cite{Son:2002sd}. See also \cite{Donos:2014gya} for a careful derivation of the gauge invariant Green's function matrix. Notice that for one field case without mass term, this is the well known structure of the retarded Green's function: $\mathbb{A}=0$ and $ G^R \sim \mathbb{O}/\mathbb{S}$. 

In summary, to compute the retarded Green's function.  We need four square matrices of order $N$(the number of fields): $\mathbb{A}, \mathbb{B}, \mathbb{S}, \mathbb{O}$. $\mathbb{A}$ and $\mathbb{B}$ can be read off from the action \eqref{sb}, after taking care of all divergences by counter terms. To construct regular matrices, $\mathbb{S}$ and $\mathbb{O}$, we solve a set of differential equations $N$ times with independent initial conditions.  The retarded Green's function is schematically $\mathbb{A} + \mathbb{B}\cdot \mathbb{O} \cdot \mathbb{S}^{-1}$. There is one subtlety in our procedure.

For our equations there is one subtlety caused by a symmetry of the system.
Solving the equations near horizon with the expansions \eqref{near1}-\eqref{near2} we find that only two of $a_x^{(I)} $, $\chi^{(I)}$, and $h_{tx}^{(I)}$ are free. Therefore, we cannot make a complete basis  to construct a general ${J^a}$.  It is due to the gauge fixing $g_{rx}=0$. However, it turns out that there is a residual gauge transformation keeping $g_{rx}=0$, which is generated by the vector field 
$\xi^\mu$ of which non-vanishing component is $\xi^x= \epsilon e^{-i\omega t} $. 
So we may add a constant vector ($\mathbb{S}_{0}^{a}$) along the residual gauge orbit. 
 \begin{equation}\label{constant1}
\mathbb{S}_{0}^{a} =( 0,  1, i\beta/\omega)^T \,, 
\end{equation} 
since  ${\cal L}_\xi g_{tx}=-i\omega r^2 \xi^x$ and  ${\cal L}_\xi \varphi = \beta \xi^x$.  Notice  that  $\mathbb{S}_{0}^{a}$   satisfies the equations of motion \eqref{eom1}-\eqref{eom4}, since the residual gauge transformation leaves the linearised equation of motion invariant. Therefore, our procedure is equivalent to formally adding a constant `solutions' of the equations to the solution set $\{\mathbb{S}_{i}^{a}\}$.

With the matrices $\mathbb{S}$ and $\mathbb{O}$, which is numerically computed, we may construct a $3 \times 3$ matrix of the retarded Green's function. We will focus on the $2 \times 2$ submatrix corresponding to $a_x^{(0)}$ and $h_{tx}^{(0)}$ in \eqref{nearb}. Since $a_x^{(0)}$ is dual to $U(1)$ current $J_x$ and $h_{tx}^{(0)}$ is dual to energy-momentum tensor $T_{tx}$ 
\begin{equation}
\label{theo}
\left(\begin{array}{cc}   G_{11} & G_{12} \\   G_{21}  &  G_{22} \end{array}\right) = 
\left(\begin{array}{cc}   G^R_{J_xJ_x} & G^R_{J_xT_{tx}} \\   G^R_{T_{tx}J_x}  &  G^R_{T_{tx}T_{tx}} \end{array}\right) 
\,.
\end{equation}
From the linear response theory,  we have the following relation between the response functions and the sources:
\begin{equation}
\label{theo}
\left(\begin{array}{c}\langle J_x \rangle \\ \langle T_{tx} \rangle \end{array}\right)=
\left(\begin{array}{cc}   G_{11} & G_{12} \\   G_{21}  &  G_{22} \end{array}\right) 
\left(\begin{array}{c}  a_x^{(0)} \\  h_{tx}^{(0)}\end{array}\right)\,.
\end{equation}
We want to relate these Green's functions  to the electric ($\sigma$), thermal ($\bar{\kappa}$), thermoelectric ($\alpha, \bar{\alpha}$)  conductivities defined as
\begin{equation}
\label{pheno}
\left(\begin{array}{c}\langle J_x \rangle \\ \langle Q_x \rangle \end{array}\right)
=
\left(\begin{array}{cc}   \sigma & \alpha T \\   \bar{\alpha} T & \bar{\kappa} T \end{array}\right)
\left(\begin{array}{c} E_x \\ - (\nabla_x T)/T\end{array}\right)~,
\end{equation}
where  $Q_x$ is the heat current, $E_x$ is an electric field and $\nabla_x T$ is a temperature gradient. 
As shown in \cite{Hartnoll:2009sz,Herzog:2009xv,Kim:2014bza}, by taking into account a diffeomorphism invariance, \eqref{pheno} can be expressed as
\begin{equation}
\label{pheno1}
\left(\begin{array}{c}\langle J_x \rangle \\ \langle T_{tx} \rangle -\mu \langle J_x \rangle \end{array}\right)
=
\left(\begin{array}{cc}   \sigma & \alpha T \\   \bar{\alpha} T & \bar{\kappa} T \end{array}\right)
\left(\begin{array}{c} i\omega  ( a_x^{(0)} + \mu  h_{tx}^{(0)})   \\   i \omega  h_{tx}^{(0)} \end{array}\right).
\end{equation}
The comparison  of \eqref{theo} and  \eqref{pheno1}  yields
\begin{equation} \label{pheno2}
\left(\begin{array}{cc}   \sigma & \alpha T \\   \bar{\alpha} T & \bar{\kappa} T \end{array}\right) =
\left(
\begin{array}{cc}
 -\frac{i G_{11}}{\omega } & \frac{i (G_{11} \mu -G_{12})}{\omega } \\
 \frac{i (G_{11} \mu -G_{21})}{\omega } & -\frac{i (G_{22} -G_{22}(\omega=0)+\mu  (-G_{12}-G_{21}+G_{11} \mu ))}{\omega } \\ 
\end{array}
\right) .
\end{equation}

\subsection{Electric/thermal/thermoelectric conductivites}

\begin{sloppypar}
 \begin{figure}[]
 \centering
   {\includegraphics[width=4.5cm]{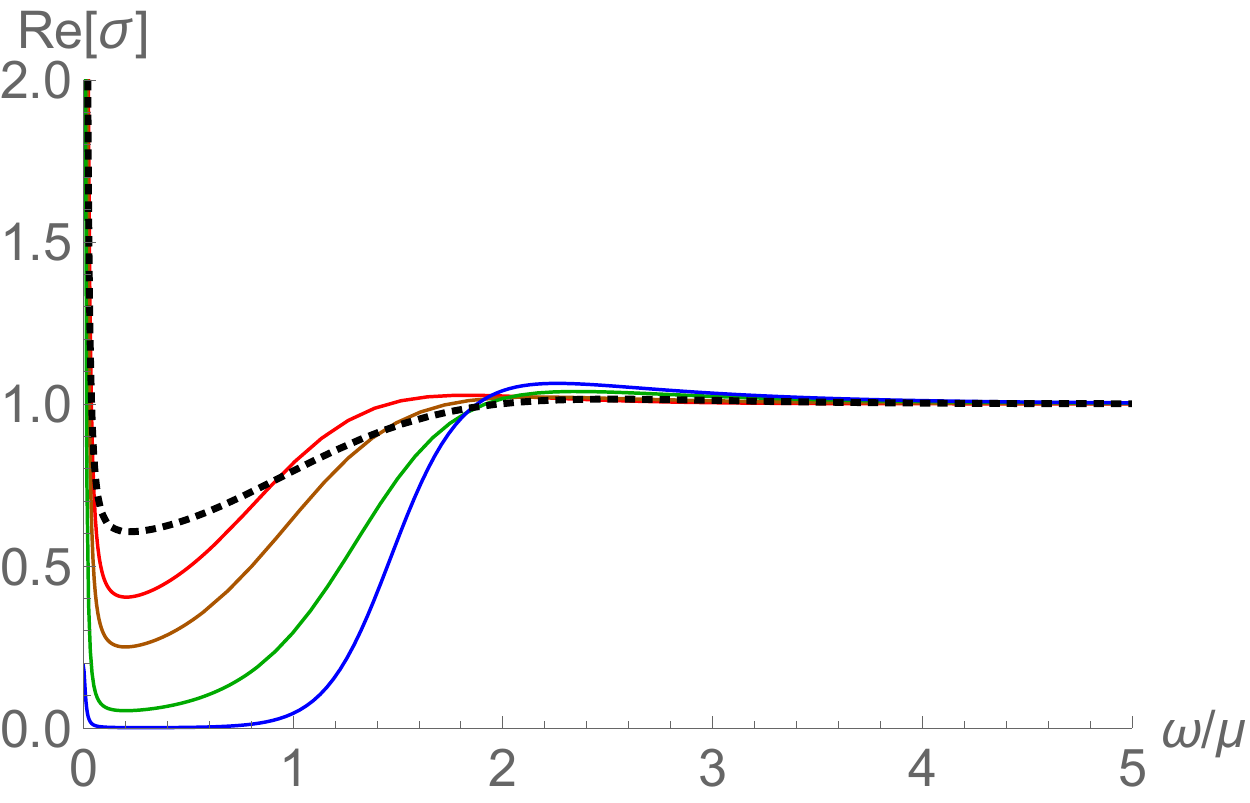} \label{}}\hspace{3mm}
   {\includegraphics[width=4.5cm]{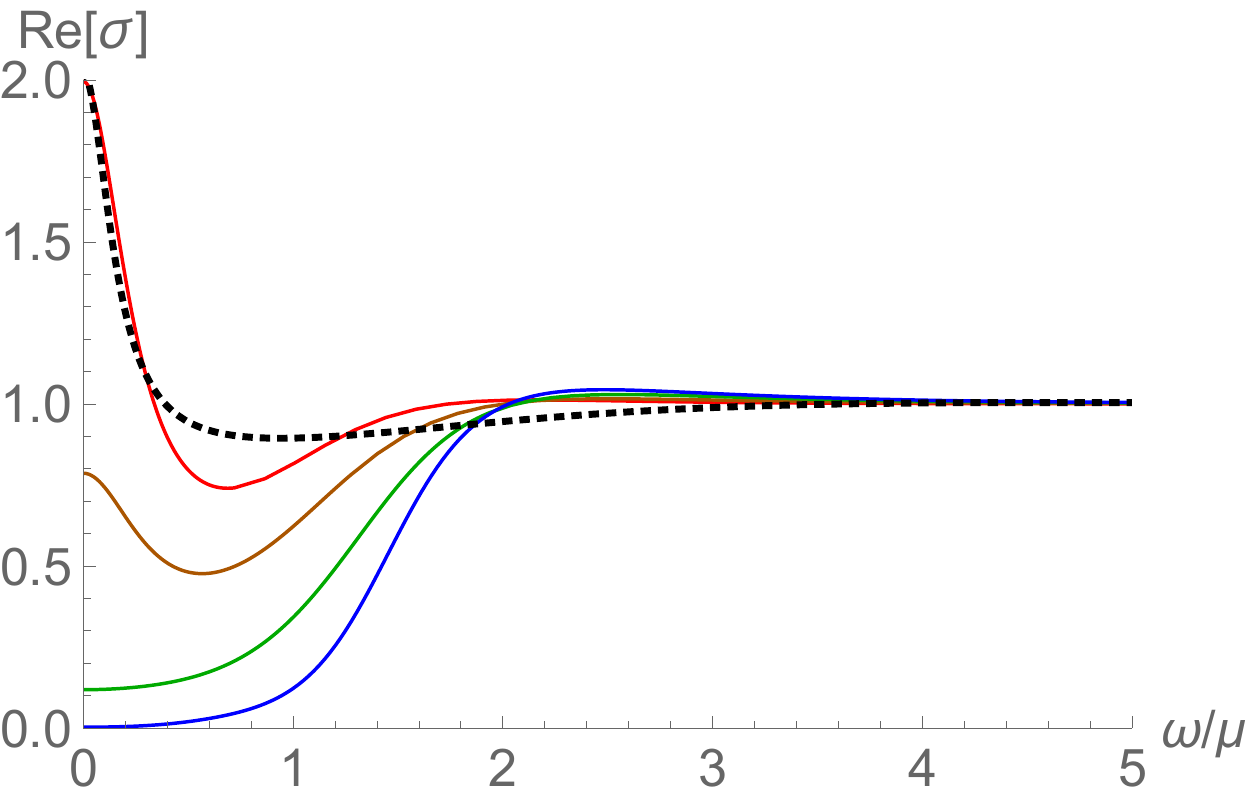} \label{}}\hspace{3mm}
   {\includegraphics[width=4.5cm]{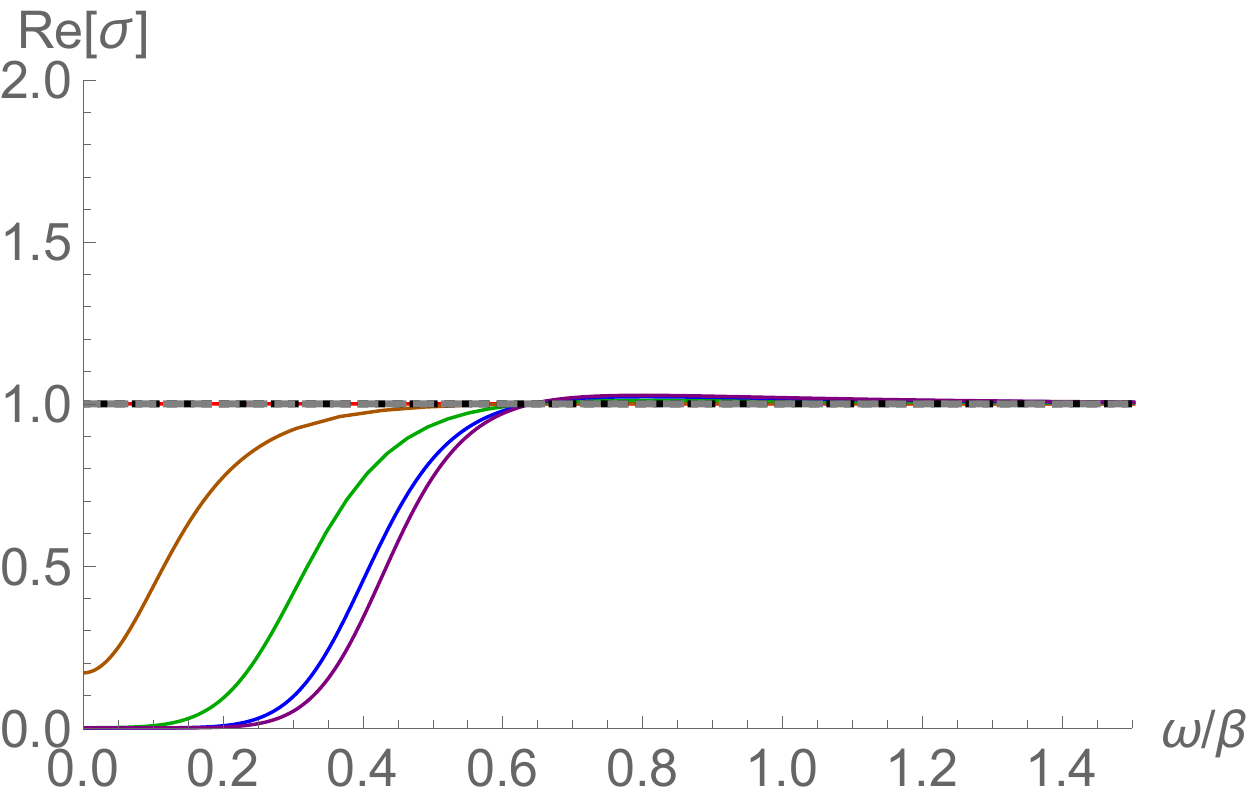} \label{}}
      \subfigure[\mbox{$ \beta/\mu = 0.1, \ \  T/T_c = 1.52,1,$} $0.94,0.76, 0.37$ (dotted, red, orange, green, blue)]
      {\includegraphics[width=4.5cm]{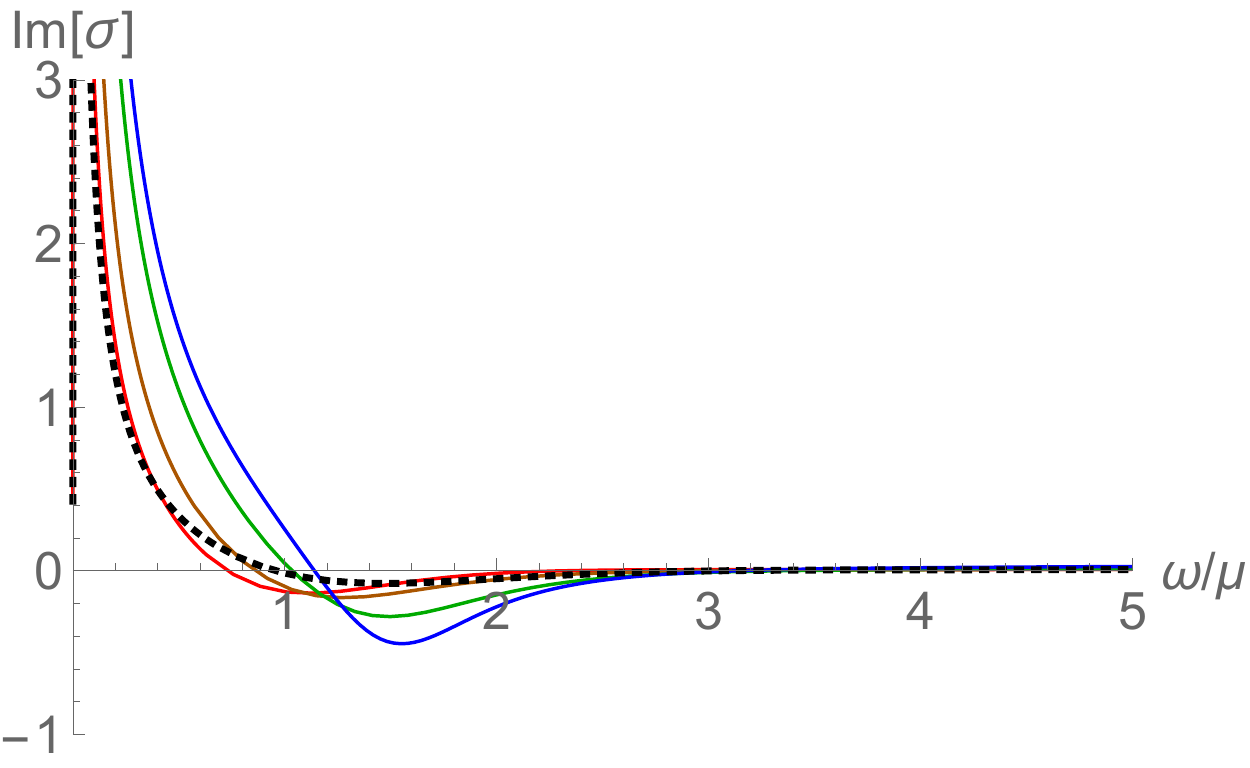} \label{}}\hspace{3mm}
     \subfigure[\mbox{$\beta/\mu = 1$, $T/T_c = 3.2, 1,0.89,$} $0.66, 0.27$ (dotted, red, orange, green, blue)]
   {\includegraphics[width=4.5cm]{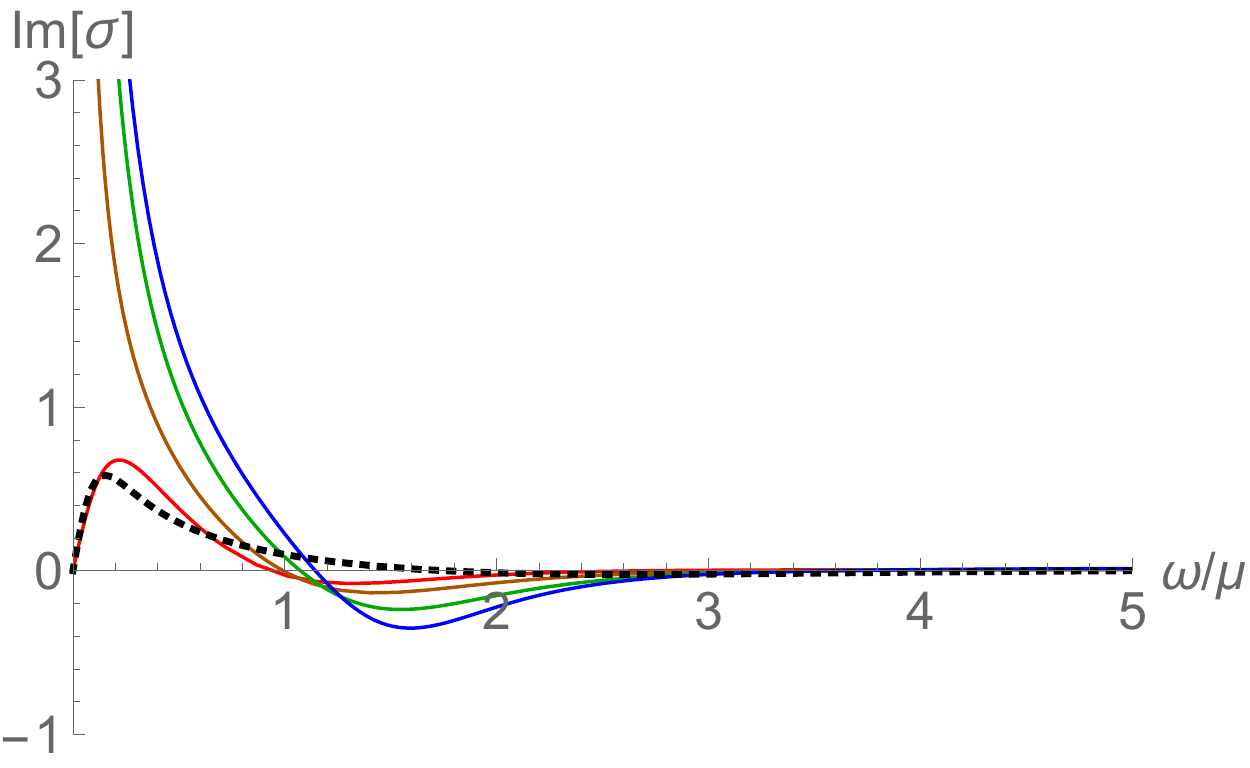} \label{}}\hspace{3mm}
     \subfigure[$\beta/\mu \rightarrow \infty  (\mu=0)$, $T/T_c = 13.2,3.5,1, 0.95,0.7,0.4,0.25 $ (dashed, dotted, red, orange, green, blue, purple)  ]
   {\includegraphics[width=4.5cm]{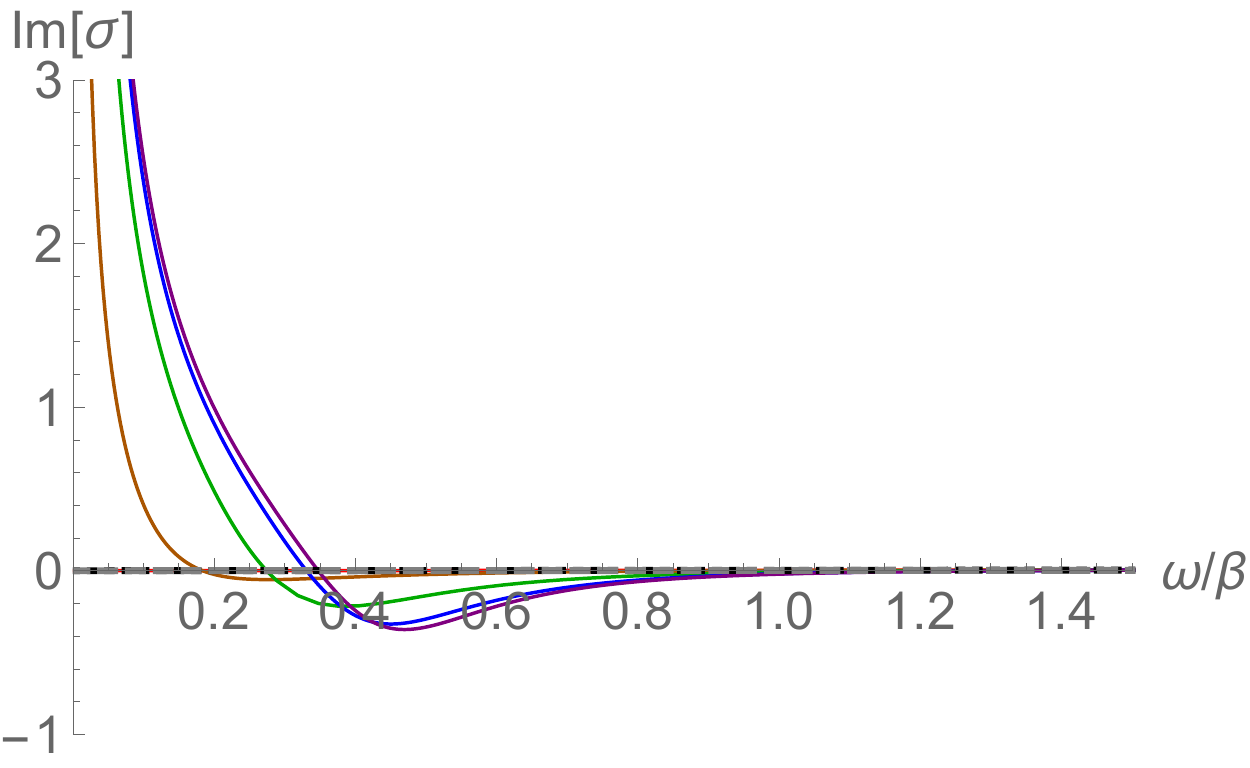} \label{}}
  \caption{ Electric conductivity($\sigma$) for three cases $\beta/\mu = 0.1, 1$ and $\infty$(or $\mu=0$) }
            \label{results1}
\end{figure}
\end{sloppypar}
\begin{figure}[]
\centering
  \subfigure[$\beta/\mu = 0.1$. Data points and fitting curves \eqref{Drude1} The purple line fits well too.]
   {\includegraphics[width=3.6cm]{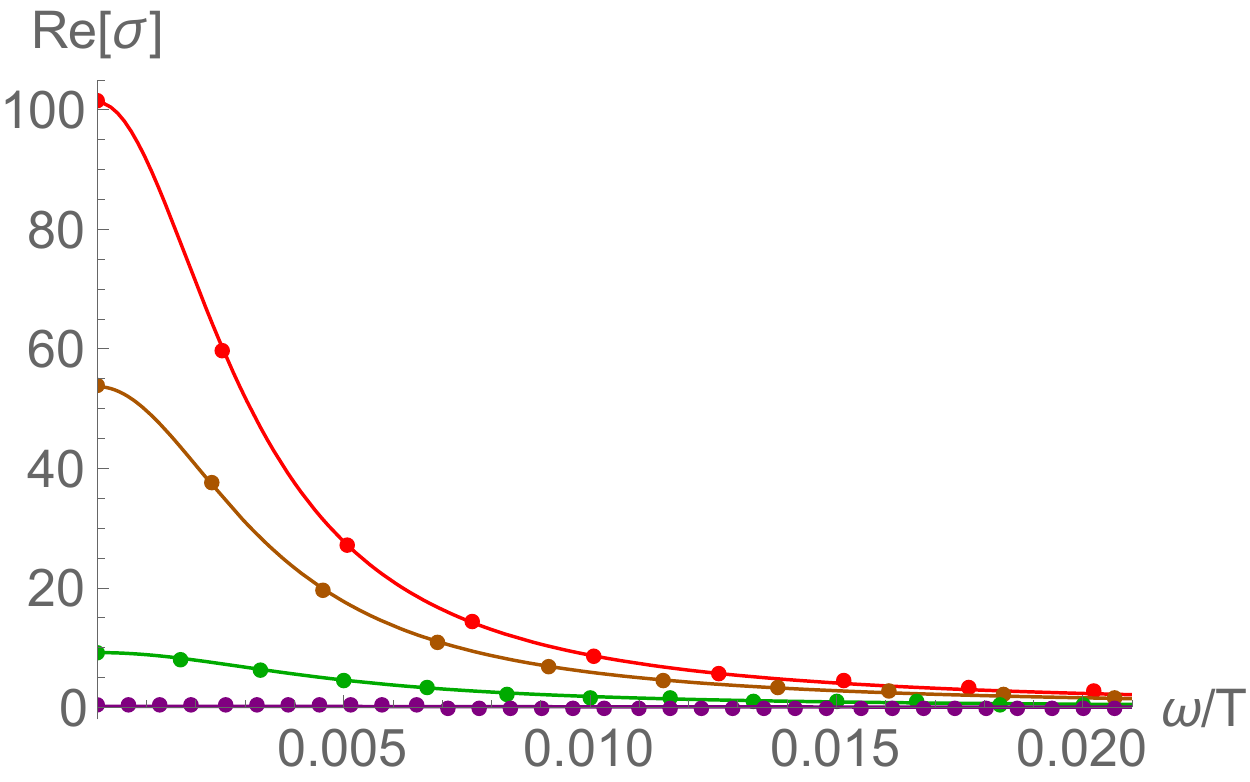} 
    \includegraphics[width=3.6cm]{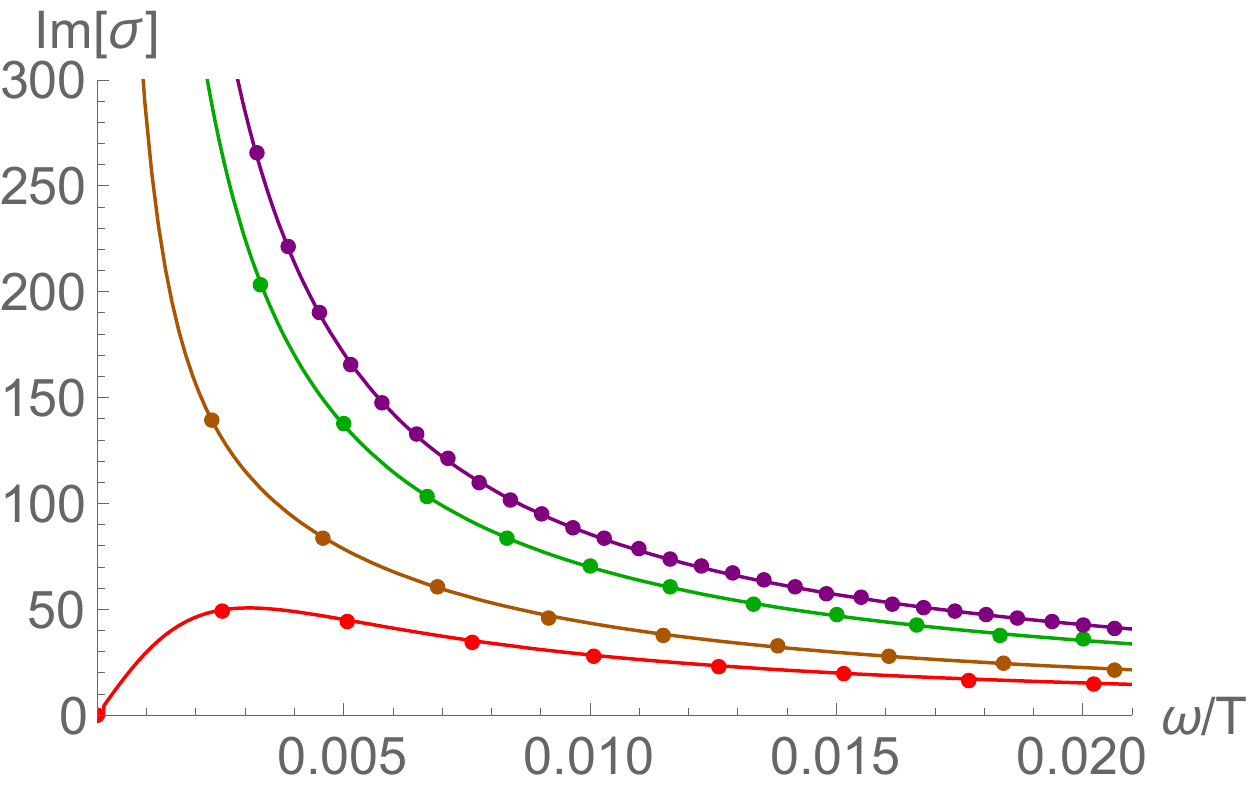}} 
     \subfigure[$\beta/\mu = 1$. Data points and fitting curves \eqref{Drude2}]
    {\includegraphics[width=3.6cm]{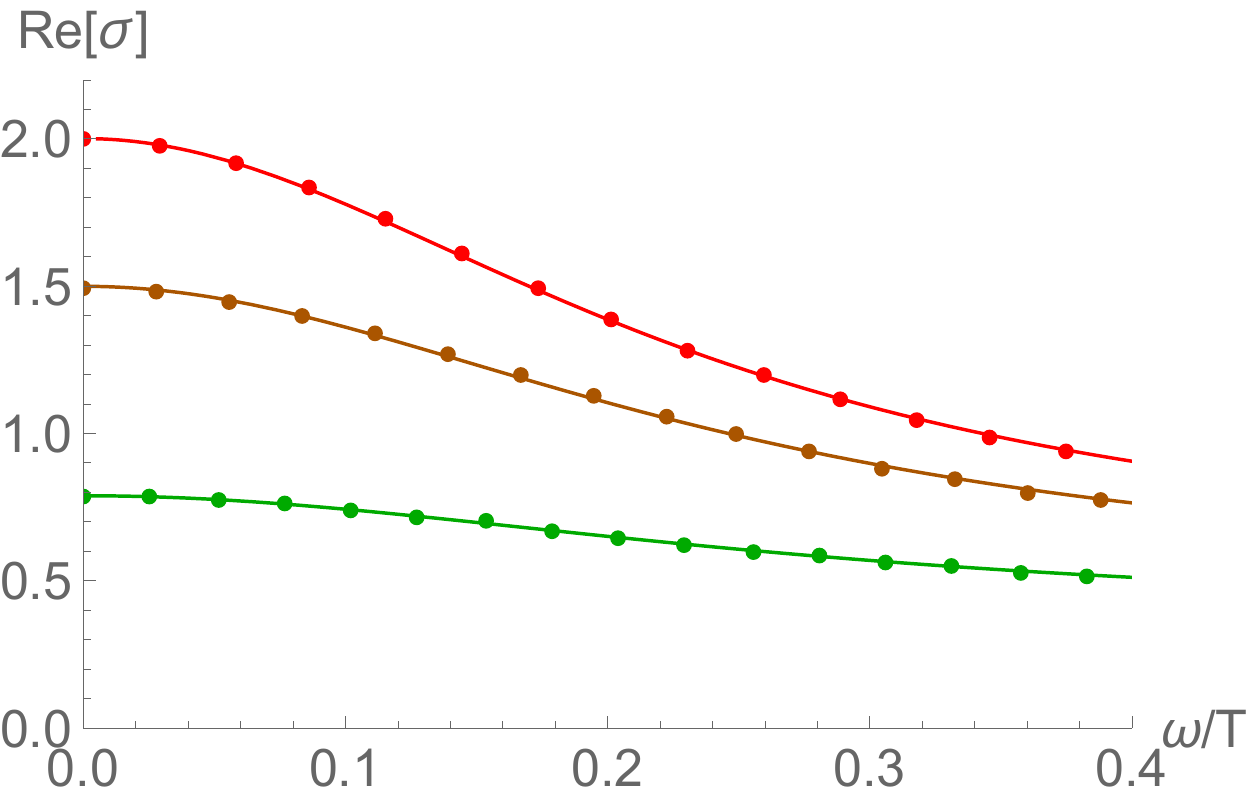}
    \includegraphics[width=3.6cm]{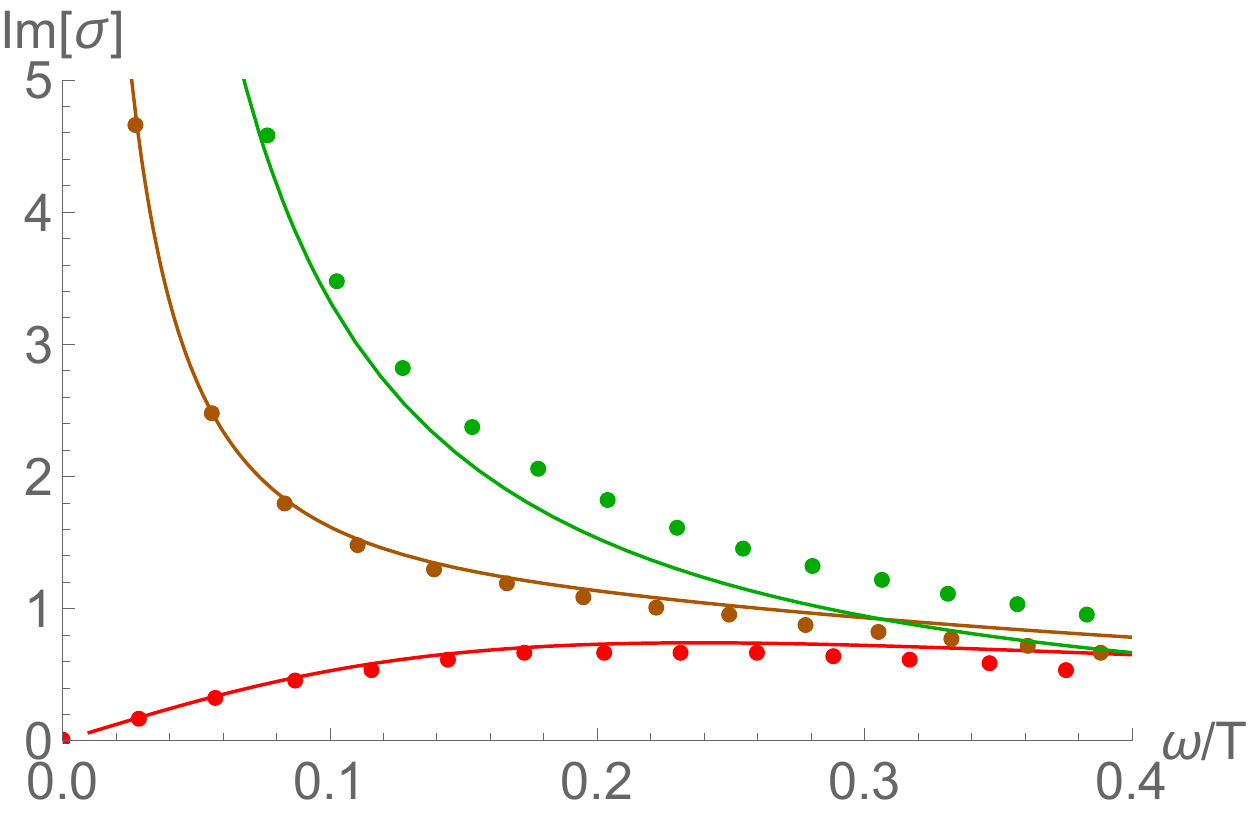}}
\caption{Near $\omega=0$ of Figure \ref{results1}(a) and (b).  $T/T_c=$ the same color as Figure \ref{results1}. Dots are the same data in Figure \ref{results1} and solid lines are Drude-like fits.}
            \label{DrudeFig}
\end{figure}

Figure \ref{results1} shows examples of electric optical conductivities($\sigma(\omega)$) for three cases of $\beta/\mu$:  $\beta/\mu = 0.1,1$ and, $\infty(\mu=0)$.  This choice of parameters also corresponds to the cases with the green($\beta/\mu = 0.1$), red($ \beta/\mu = 1$), and blue($\beta/\mu = \infty $) lines in Figure \ref{phase3D}(a) and \ref{Condensate}. The color of curves represents temperature ratio,  $T/T_c$, where $T_c$ is the critical temperature. The numerical values of temperature ratio are shown in the caption.  In particular the dotted black curve\footnote{There is also a dashed grey curve in (c) at $\mu=0$. It is not distinguishable from the red and  dotted black curves in Figure \ref{results1} and \ref{results2}, but distinguishable in Figure \ref{results3}. } is for the temperature above $T_c$, which is in metal phase and the red curve corresponds to the critical temperature (in practice, it is slightly higher than the transition temperature.). The first row shows the real part of electric conductivity (Re[$\sigma$]) and the second row shows the imaginary part of electric conductivity (Im[$\sigma$]).   

One common important feature in Figure \ref{results1} is the appearance of $1/\omega$ pole in Im[$\sigma$] below the critical temperature, while the disappearance of  $1/\omega$ pole above the critical temperature.  By the Kramers-Kronig relation, $1/\omega$ pole in Im[$\sigma$]  implies the delta function at $\omega=0$ in Re[$\sigma$].  Therefore, in normal phase the DC conductivity is finite due to momentum relaxation and in superconducting phase the DC conductivity is infinite, which is one of the hallmarks of superconductor.

Roughly at $T/T_c > 0.5$, in addition to the delta function at $\omega=0$,  there is still a finite value of DC  Re[$\sigma$] in superconducting phase. It may be interpreted as a contribution from normal components in superconducting phase, implying a two-fluid model.  For small $\omega$ there is a Drude-like peak in some cases in (a) and (b) of Figure \ref{results1}. For smaller $\beta/\mu$ or at higher temperature, the peak becomes sharper. 
For the sake of comparison we used a similar scales in (a),(b) and (c) of Figure \ref{results1}, which hides the structure of (a) near $\omega = 0$. Therefore we zoom in Figure  \ref{results1}(a)  in Figure \ref{DrudeFig}(a). The data points well fit to 
the formula (solid lines)
\begin{equation} \label{Drude1}
\sigma(\bar{\omega}) = i \frac{K_s}{\bar{\omega}} + \frac{K_n \tau}{1-i \bar{\omega} \tau} \,,
\end{equation}
where $\bar{\omega} \equiv \omega/\mu$ and $K_s$ and $K_n$ are supposed to be proportional to the superfluid density and normal fluid density. 
For $\beta/\mu = 1$  the formula \eqref{Drude1} does not work and the data(red, orange, green) better fit to 
\begin{equation} \label{Drude2}
\sigma(\bar{\omega}) = i \frac{K_s}{\bar{\omega}} + \frac{K_n \tau}{1-i \bar{\omega} \tau} + K_0 \,,
\end{equation}
which is shown in Figure \ref{DrudeFig}(b).  $K_0$ is related to pair creation and it was necessary also in metal phase. The existence of $K_0$ is most apparent in Figure \ref{results1}(c), where $\mu = K_n = 0$.   Indeed \eqref{Drude1} is understood as an approximation of \eqref{Drude2} when $K_0$ is negligible compared to $K_n \tau$. 
As temperature is lowered $K_n$ and $K_0$ is reduced while $K_s$ is enhanced. $K_0$ becomes zero at low temperature (green and blue line in Figure \ref{results1}(c)), but it is possible that $K_n$ is finite even at zero $T$ \cite{Horowitz:2013jaa,Zeng:2014uoa,Ling:2014laa}. 

\begin{figure}[]
 \centering{\includegraphics[scale=0.55]{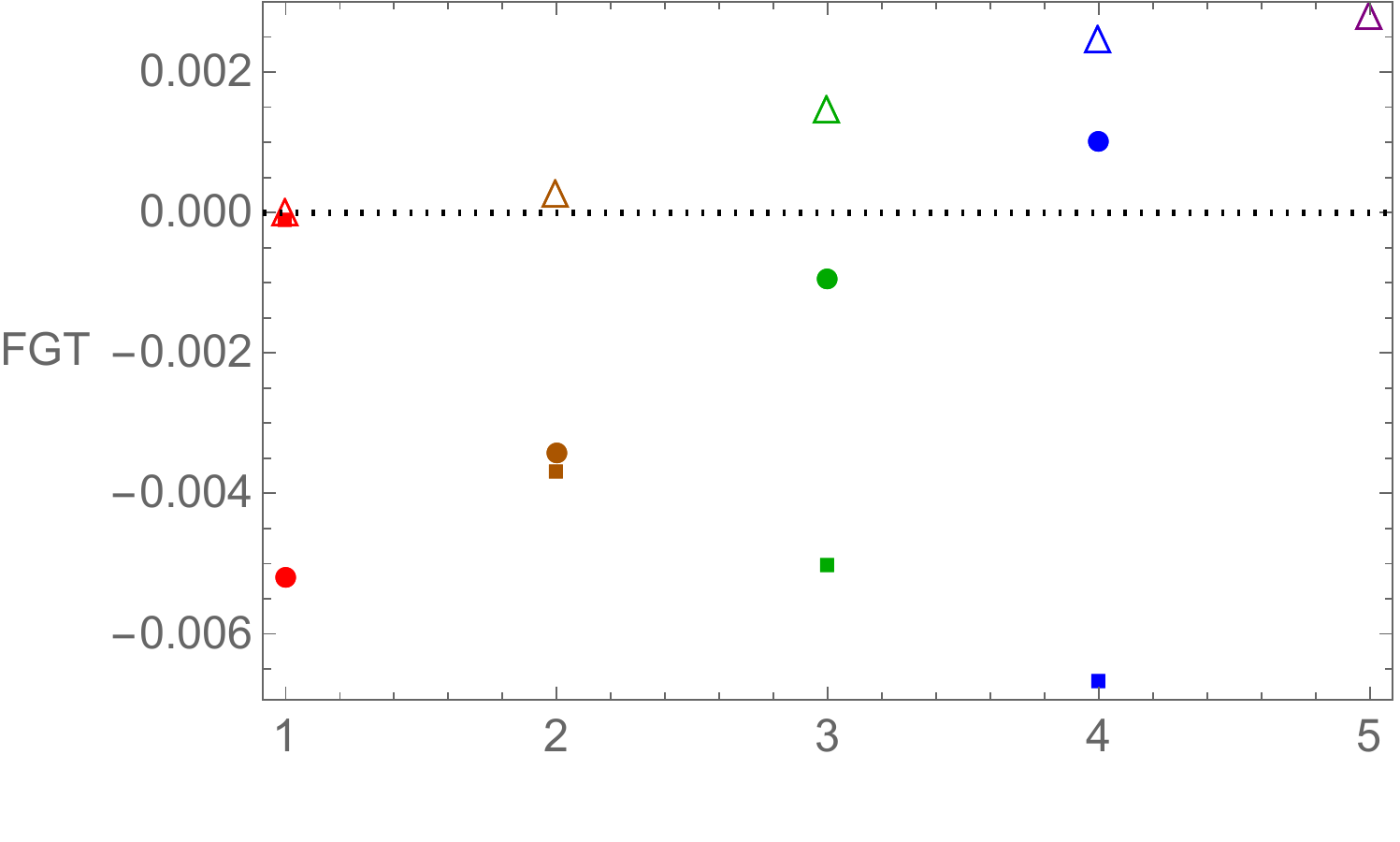}}
\caption{The numerical errors of the FGT sum rule. FGT is defined in \eqref{FGTeq}. Data for $\frac{\beta}{\mu}=0.1$, $\frac{\beta}{\mu}=1$, and $\frac{\beta}{\mu}=\infty$ are plotted as square, circle, and triangle respectively, and colors indicates the temperatures with the same colors in Figure \ref{results1}.  }
\label{FGT}
\end{figure}

If $\beta \gg \mu$, it was shown that the coherent metal phase becomes incoherent, where the Drude peak becomes a non-Drude peak~\cite{Kim:2014bza}.  The Figure 5 in \cite{Kim:2014bza} suggests that the critical $\beta/\mu$ is around $1/2$ in metal phase.  Therefore, it is suspected that, If $\beta > (1/2)\mu$, \eqref{Drude2} does not work in superconductor phase either.  Indeed we see this is the case.  When $\beta = \mu$: the fit of Figure \ref{DrudeFig} (b) is not as good as (a) and starts deviating from \eqref{Drude2}\footnote{The Drude nature  in superconducting phase will be related to $K_n$ rather than $\mu$. For a better understanding of the range of applicability of the Drude model, it is important to analyse $K_n$ more extensively.}.

\begin{sloppypar}
 \begin{figure}[]
 \centering
     {\includegraphics[width=4.5cm]{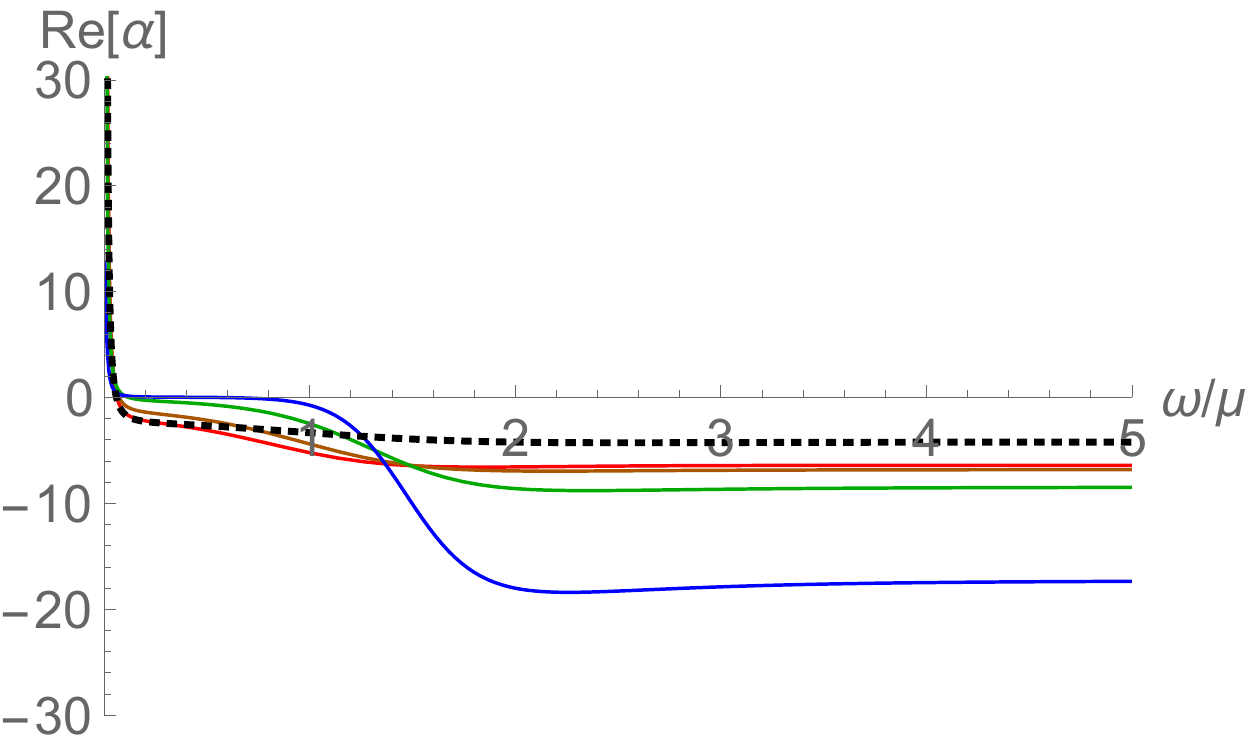} \label{}}\hspace{3mm}
   {\includegraphics[width=4.5cm]{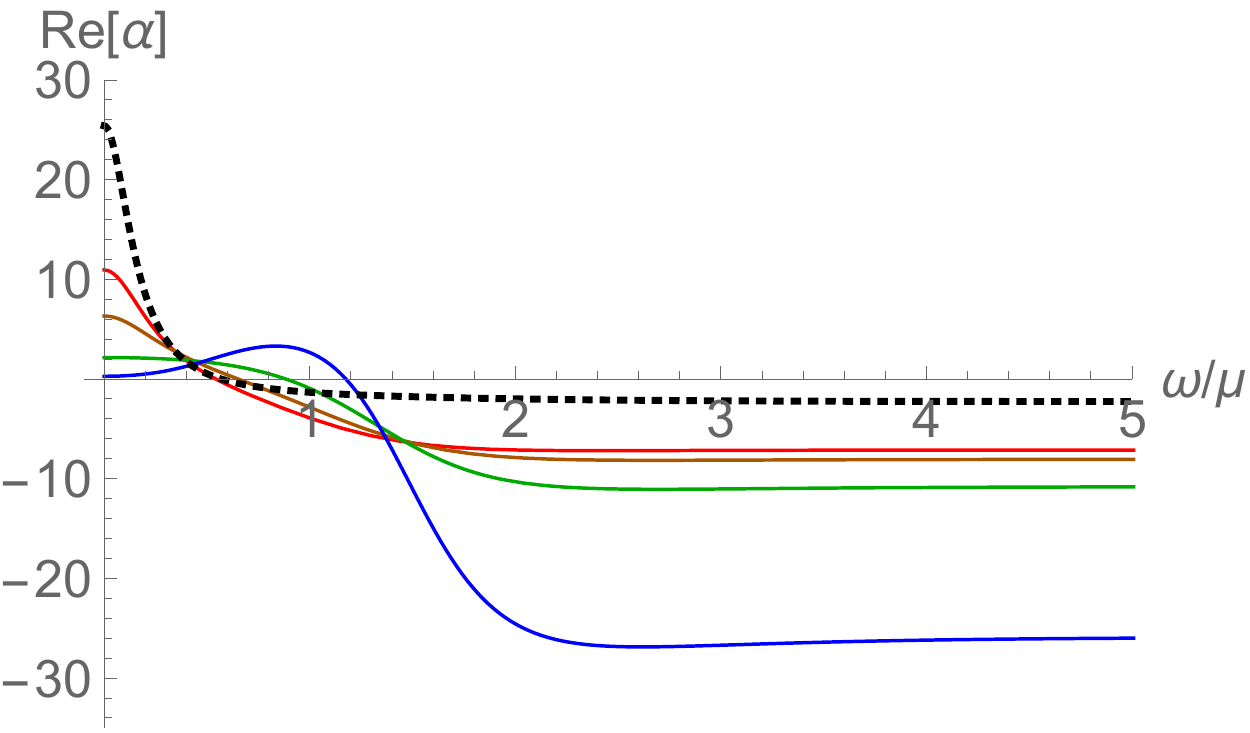} \label{}}\hspace{3mm}
   {\includegraphics[width=4.5cm]{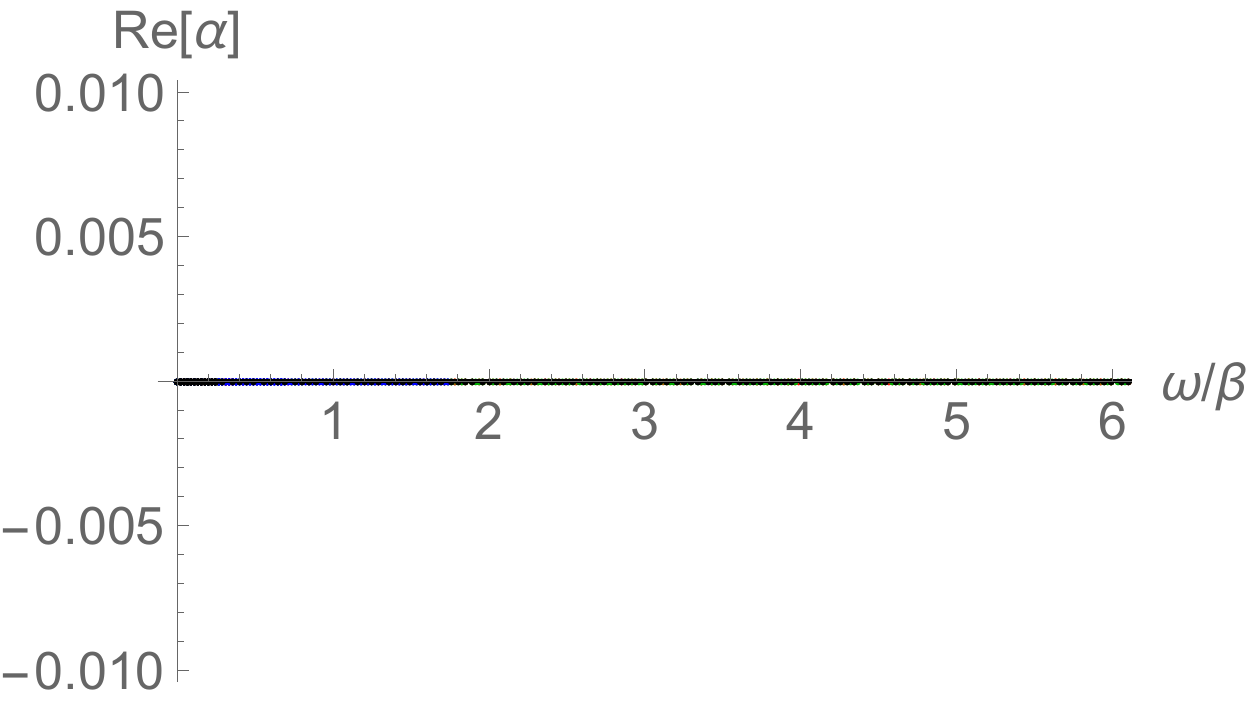} \label{}}
      \subfigure[\mbox{$ \beta/\mu = 0.1, \ \  T/T_c = 1.52,1,$} $0.94,0.76, 0.37$ (dotted, red, orange, green, blue)]
      {\includegraphics[width=4.5cm]{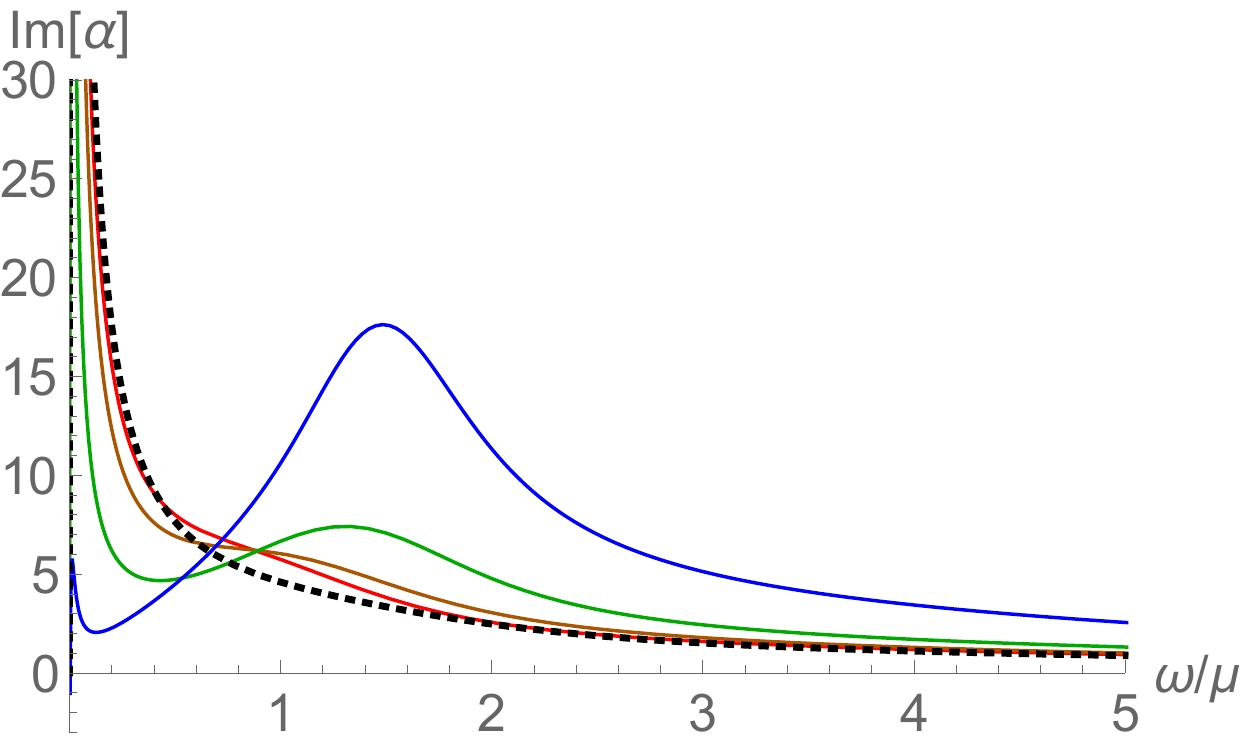} \label{}}\hspace{3mm}
    \subfigure[\mbox{$\beta/\mu = 1$, $T/T_c = 3.2, 1,0.89,$} $0.66, 0.27$ (dotted, red, orange, green, blue)]
   {\includegraphics[width=4.5cm]{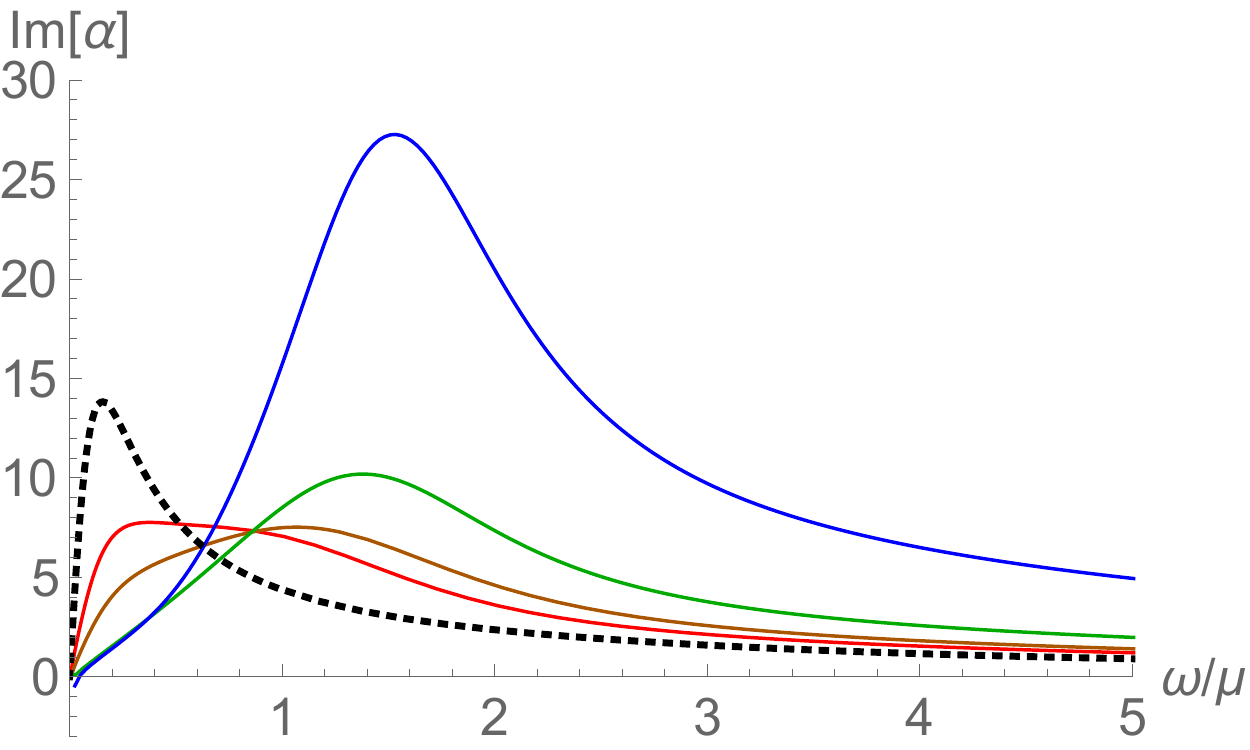} \label{}}\hspace{3mm}
     \subfigure[$\beta/\mu \rightarrow \infty  (\mu=0)$, $T/T_c = 13.2,3.5,1, 0.95,0.7,0.4,0.25 $ (dashed, dotted, red, orange, green, blue, purple)  ]
       {\includegraphics[width=4.5cm]{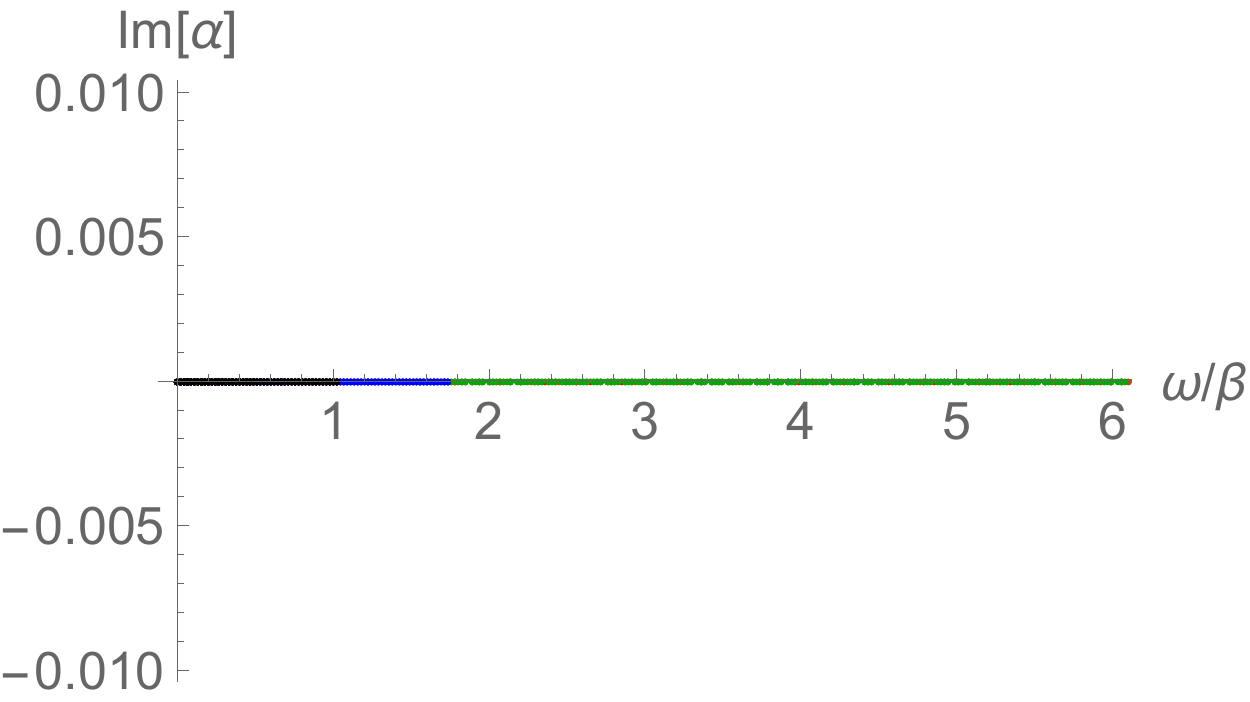} \label{}}
  \caption{ Thermoelectric conductivity($\alpha$) for three cases $\beta/\mu = 0.1, 1$ and $\infty$(or $\mu=0$) } 
            \label{results2}
\end{figure}
\end{sloppypar}

\begin{sloppypar}
 \begin{figure}[]
 \centering
     {\includegraphics[width=4.5cm]{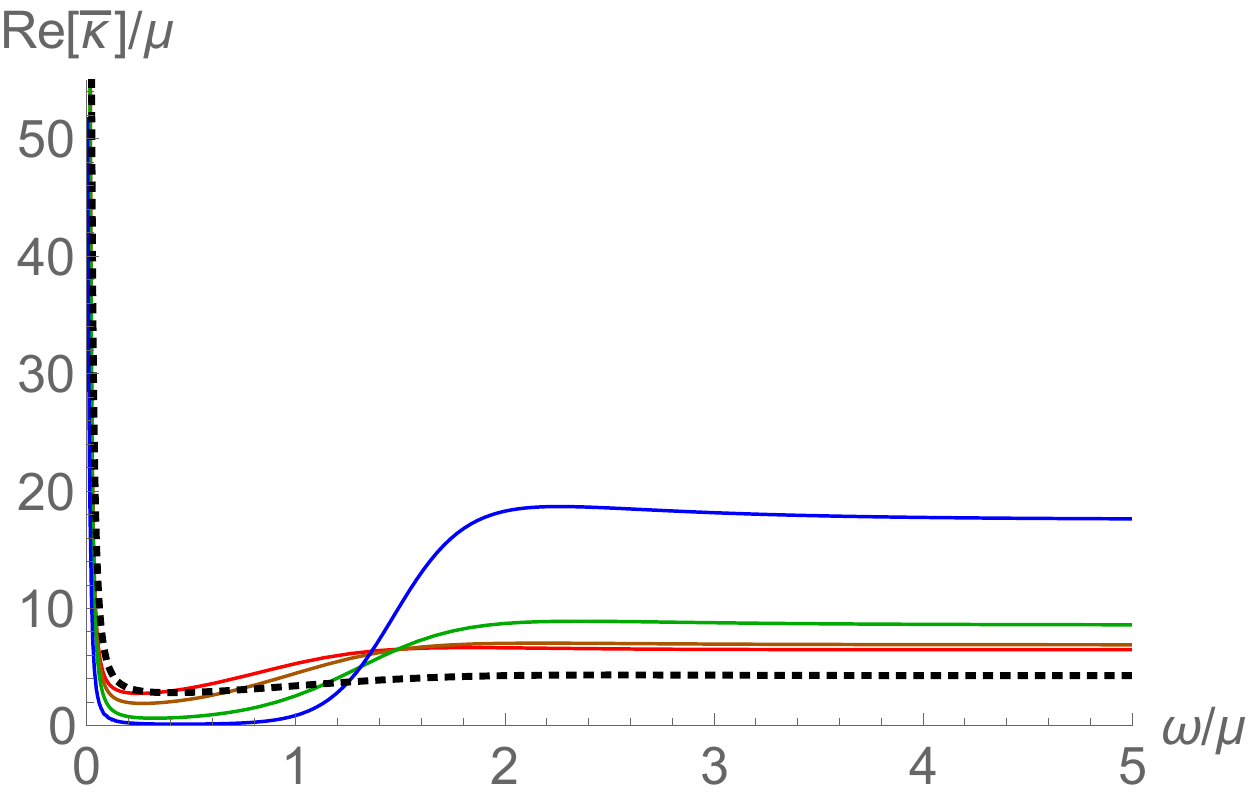} \label{}}\hspace{3mm}
   {\includegraphics[width=4.5cm]{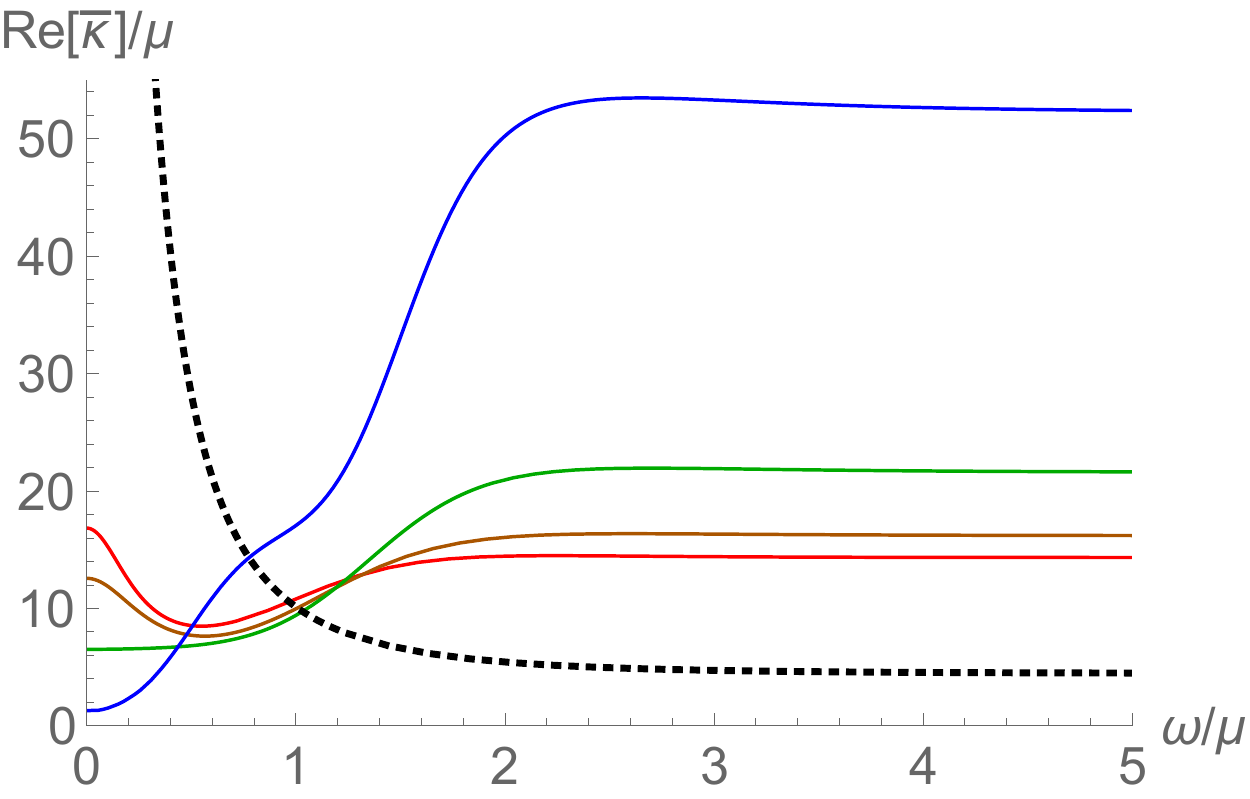} \label{}}\hspace{3mm}
   {\includegraphics[width=4.5cm]{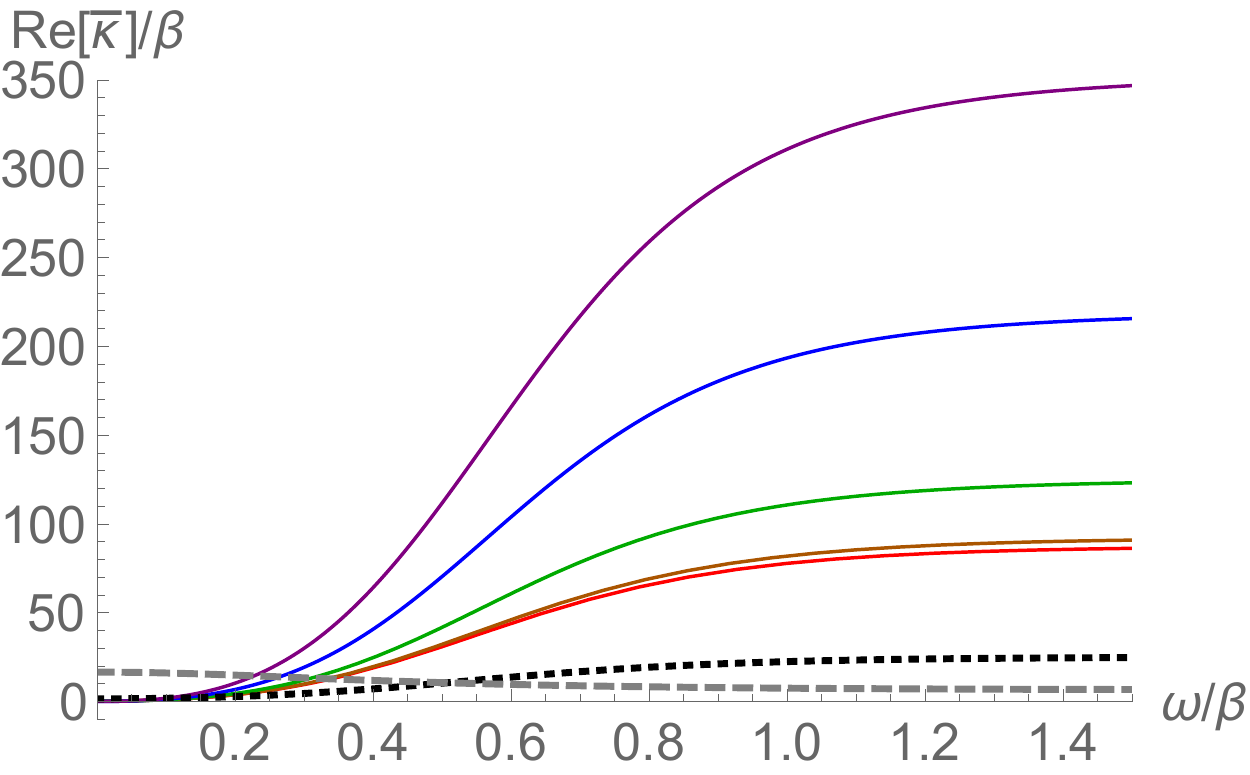} \label{}}
      \subfigure[\mbox{$ \beta/\mu = 0.1, \ \  T/T_c = 1.52,1,$} $0.94,0.76, 0.37$ (dotted, red, orange, green, blue)]
 {\includegraphics[width=4.5cm]{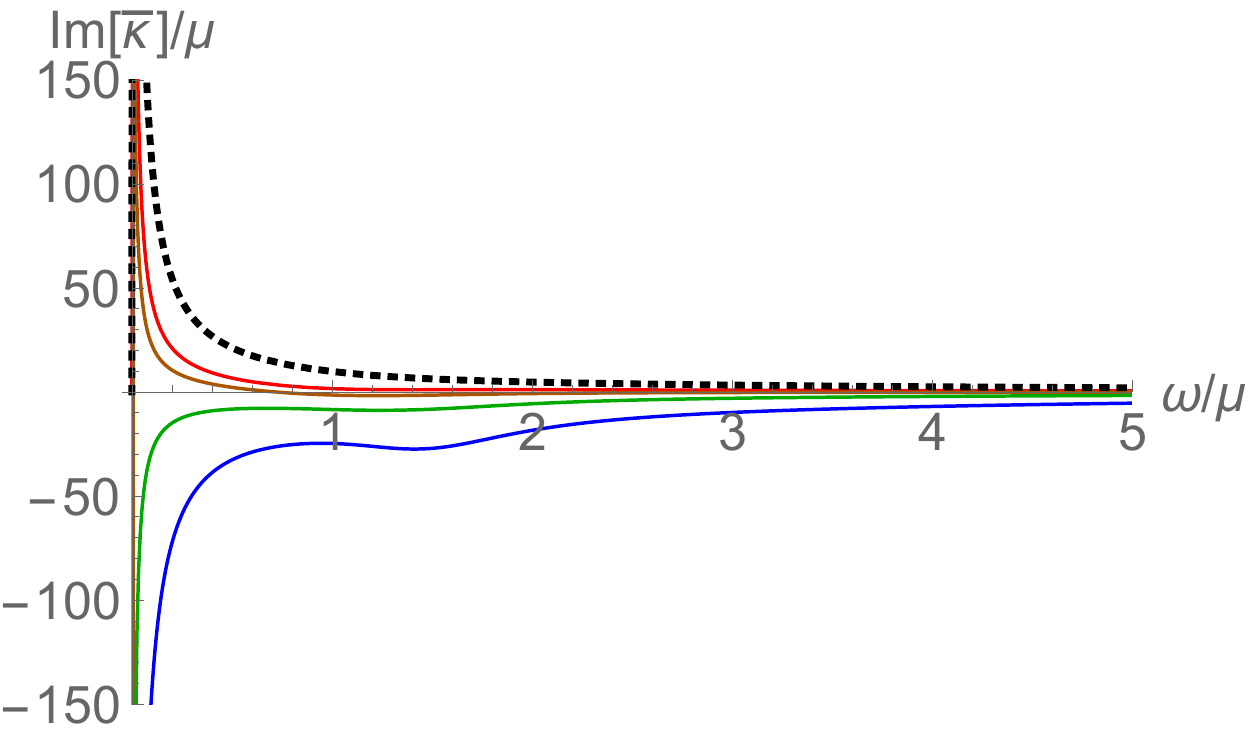} \label{}}\hspace{3mm}
    \subfigure[\mbox{$\beta/\mu = 1$, $T/T_c = 3.2, 1,0.89,$} $0.66, 0.27$ (dotted, red, orange, green, blue)]
   {\includegraphics[width=4.5cm]{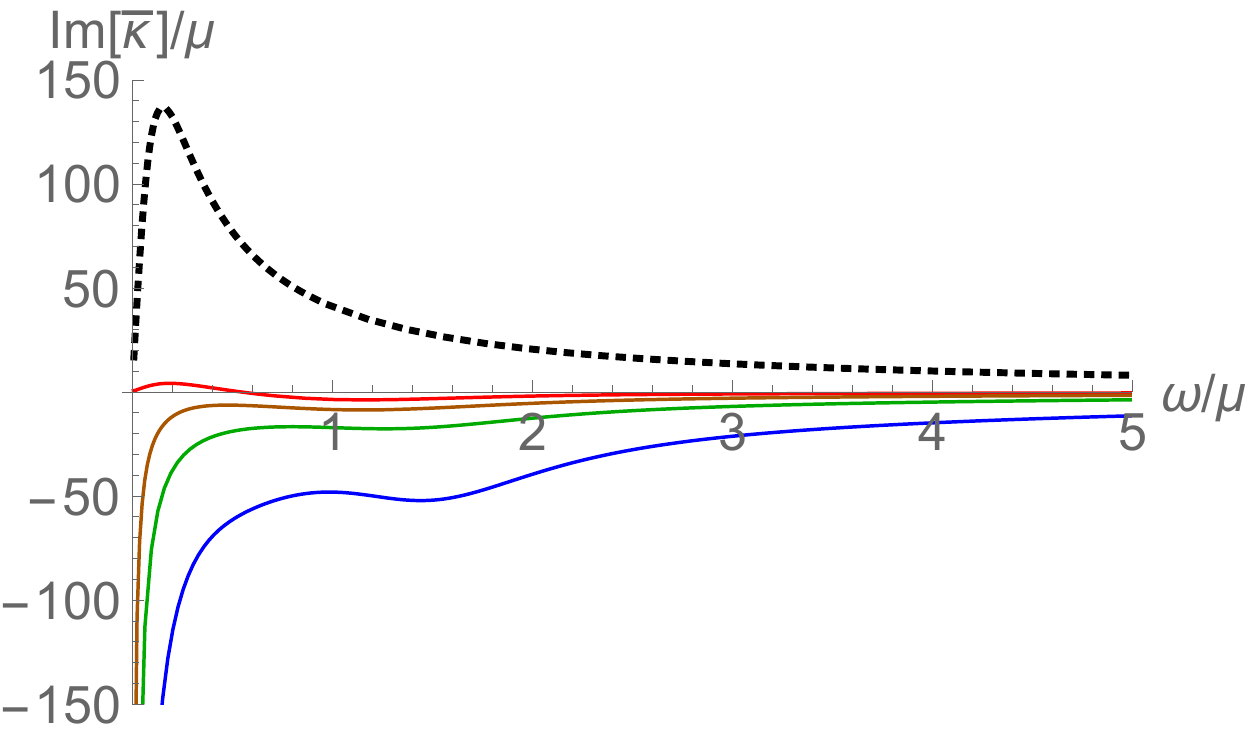} \label{}}\hspace{3mm}
     \subfigure[$\beta/\mu \rightarrow \infty  (\mu=0)$, $T/T_c = 13.2,3.5,1, 0.95,0.7,0.4,0.25 $ (dashed, dotted, red, orange, green, blue, purple)  ]
   {\includegraphics[width=4.5cm]{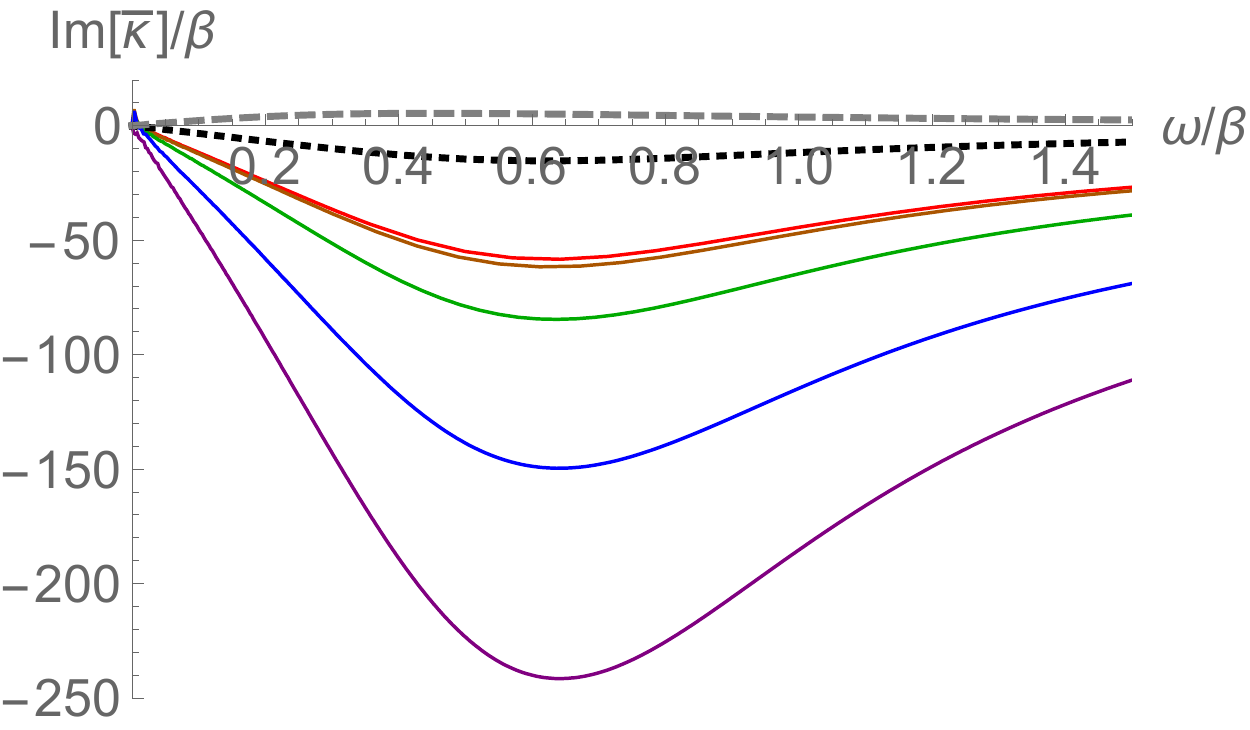} \label{}}
  \caption{ Thermal conductivity($\bar{\kappa}$) for three cases $\beta/\mu = 0.1, 1$ and $\infty$(or $\mu=0$) } 
            \label{results3}
\end{figure}
\end{sloppypar}

As temperature  goes down ($T<T_c$) the spectral weight of Re[$\sigma$]  is reduced while  $K_s$ of Im[$\sigma$] is enhanced. This transfer of the spectral weight to $K_s$ may be quantified by the Ferrell-Glover-Tinkham (FGT) sum rule:
\begin{equation} \label{FGTeq}
\mathrm{FGT} \equiv \int^{\infty}_{0^+} \dd \omega \mathrm{Re} [\sigma_n (\omega) - \sigma_s (\omega)] - \frac{\pi}{2} K_s =0 \,,
\end{equation}
where $\sigma_s$($\sigma_n$) is the conductivity at $T<T_c$($T>T_c$).  $\sigma_n$ can be taken for any temperature for $T>T_c$ since the spectral weight is constant in metal phase ~\cite{Kim:2014bza}.  
We computed \eqref{FGTeq} numerically for all cases in Figure \ref{results1} and showed that  the FGT sum rule is satisfied up to $10^{-3}$ in  Figure \ref{FGT}.

At $\mu=0$ and $\beta \ne 0$ (Figure \ref{results1} (c)) there is no net charge and no Drude peak. The plots are very similar to the case $\beta = 0$ at finite $\mu$.  For example, see Figure 6 in \cite{Hartnoll:2009sz} or Figure 1 in \cite{Kim:2014bza}, where the infinite
DC conductivity ($1/\omega$ pole in the imaginary part) is due to translation invariance with finite $\mu$. Here, there is no $\mu$ and no translation invariance. So the delta function must have a different origin, which may be a new type of superconductivity.  Interestingly, even in this case, the FGT the sum rule works. The deficit of spectral function may be interpreted as a deficit of `particle-anti-particle pairs', which will condense.  
It may imply a new `pairing mechanism' of particles and anti particles interacting with $\beta$, which may be interpreted as  kind of `impurity'~\cite{Kim:2014bza}.

\begin{figure}[]
\centering
  \subfigure[$\alpha$, $\beta/\mu = 0.1$, Figure \ref{results2}(a)]
   {\includegraphics[width=3.6cm]{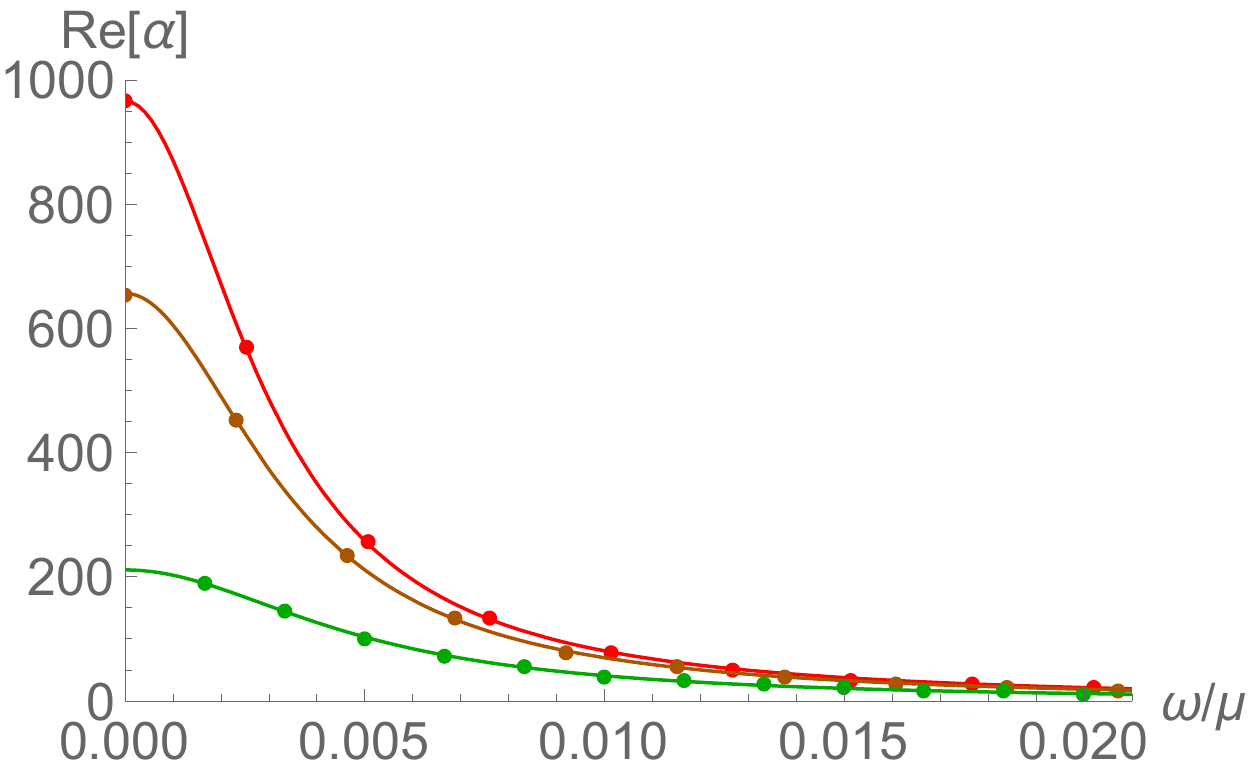} 
    \includegraphics[width=3.6cm]{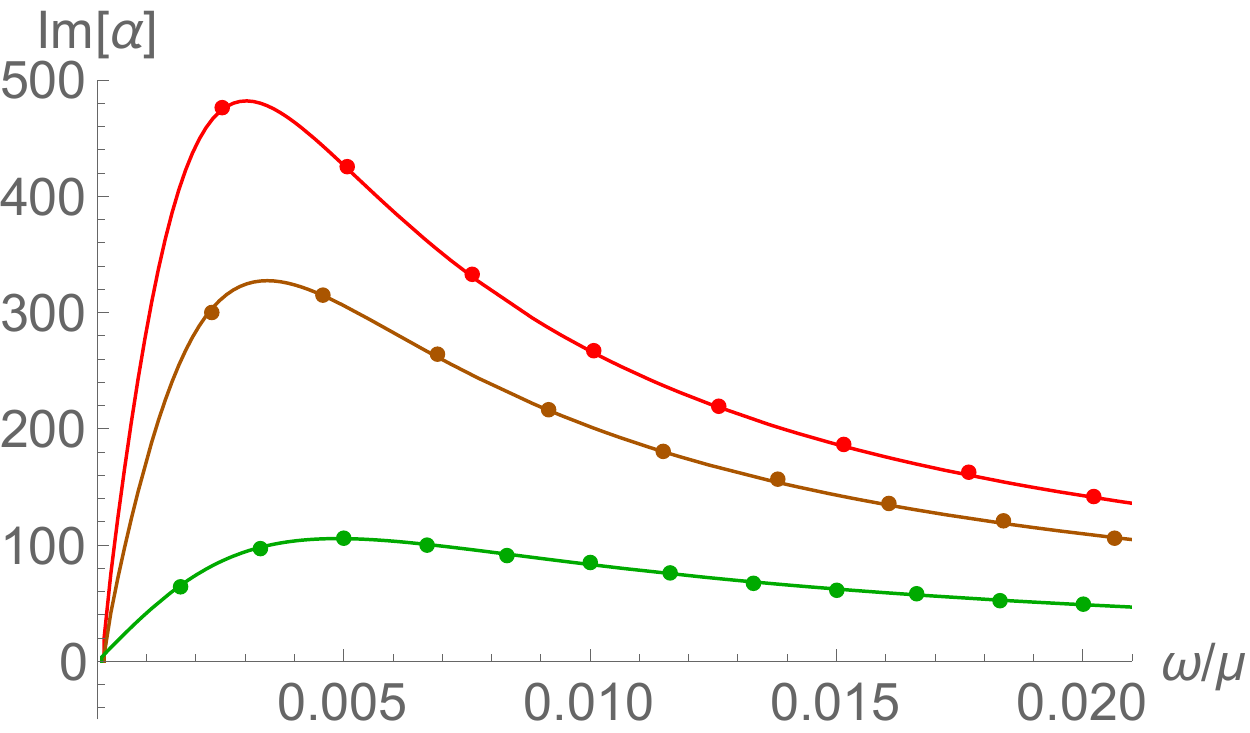}} 
     \subfigure[$\bar{\kappa}$, $\beta/\mu = 0.1$, Figure \ref{results3}(a)]
    {\includegraphics[width=3.6cm]{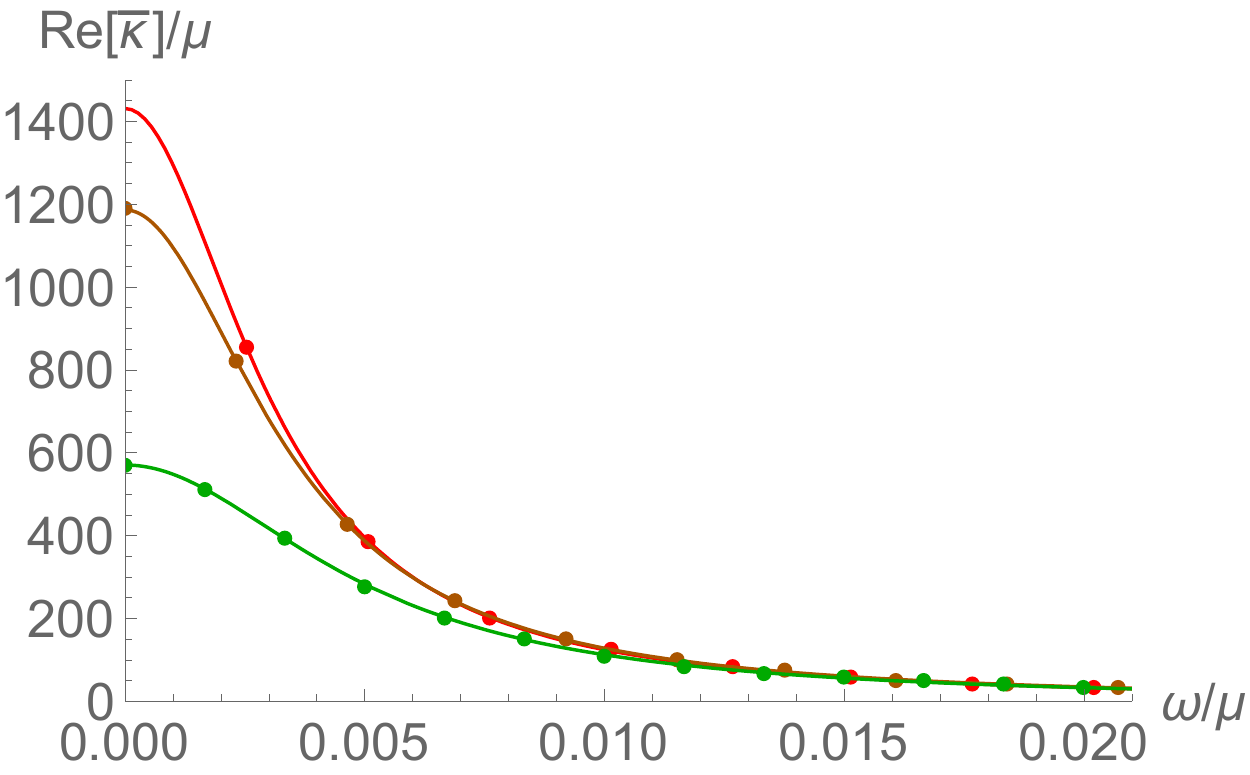}
    \includegraphics[width=3.6cm]{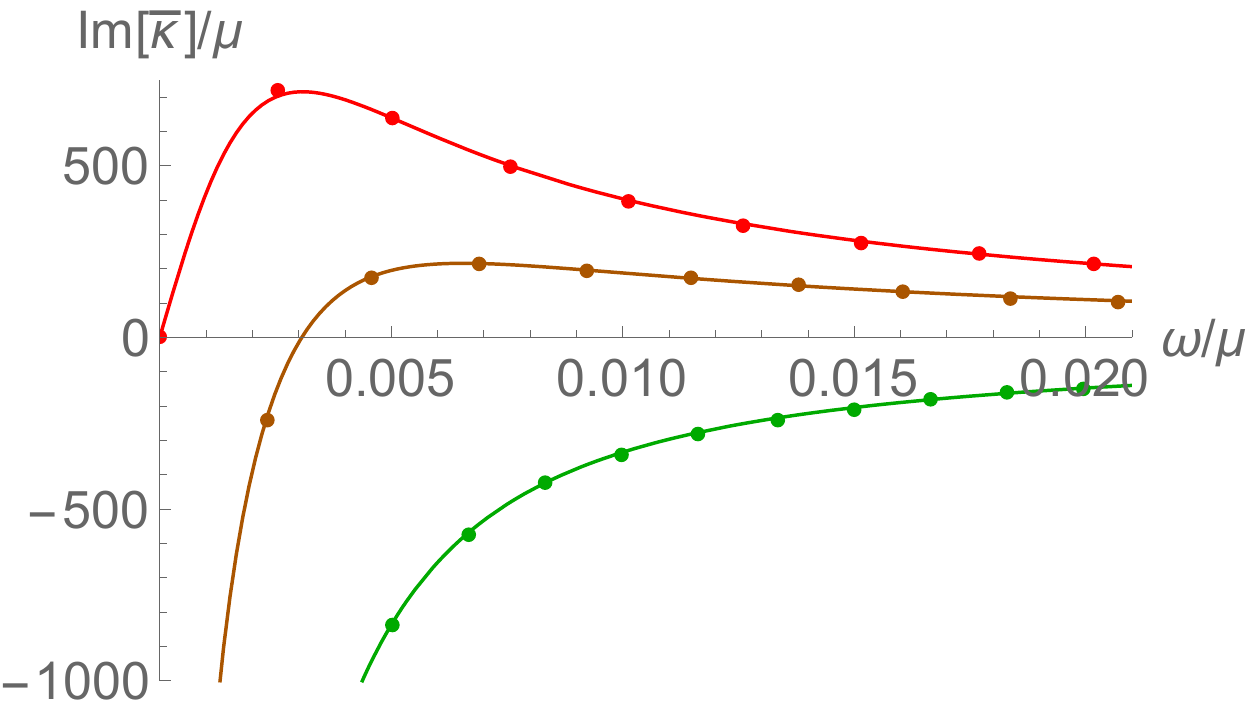}}
\caption{Near $\omega=0$.  Dots are the same data in Figure \ref{results2} and \ref{results3} and solid lines are  \eqref{Drude1}. Colors represent the same temperature.}
            \label{DrudeFig2}
\end{figure}

\begin{figure}[]
\centering
   \includegraphics[width=5cm]{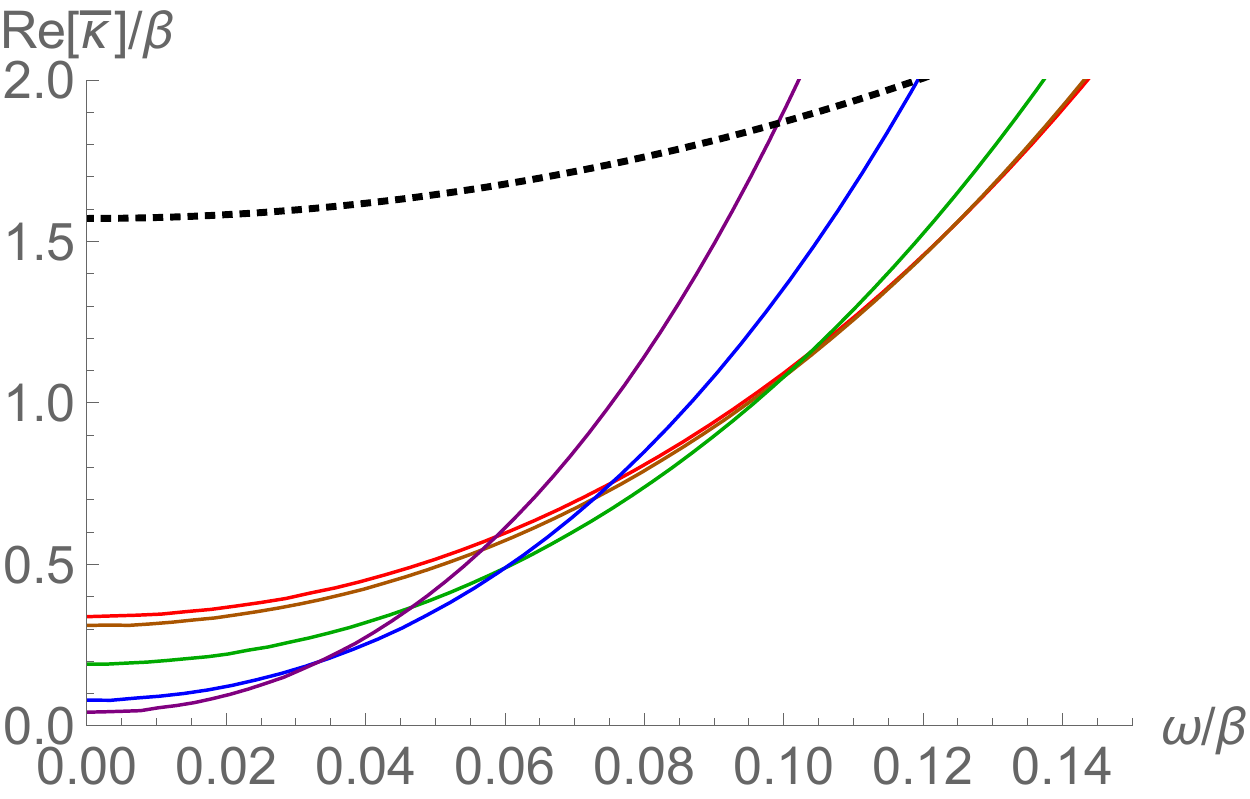} 
\caption{Re$[\bar{\kappa}]/\beta$ near $\omega=0$ at $\mu=0$, Figure  \ref{results3} (c)}
            \label{kappazoomFig}
\end{figure}

Figure \ref{results2} and \ref{results3}   show the thermoelectric conductivity ($\alpha$) and  the  thermal conductivity ($\bar{\kappa}$) for the same parameters as Figure \ref{results1}. At finite $\mu$ ((a) and (b)) the qualitative structure of Re[$\alpha$] and Re[$\bar{\kappa}$] at small $\omega$ are similar to Re[$\sigma$].  For the sake of comparison we used a similar scales in (a),(b). To see the structure of (a) near $\omega = 0$ we zoom in Figure  \ref{results2}(a) and \ref{results3}(a) in Figure \ref{DrudeFig2}. The data points fit to \eqref{Drude1}.  At $\mu=0$,  the DC values of Re[$\bar{\kappa}$] decreases quickly as temperature goes down as shown in Figure \ref{kappazoomFig}.  The thermoelectric conductivity  $\alpha$ vanishes, if $\mu=0$, which is due to particle-hole symmetry.     As $\omega \rightarrow \infty$, $\alpha$ and $\bar{\kappa}$ approaches to 
\begin{equation}
\alpha \rightarrow - \frac{\mu}{T}\,,  \qquad  \bar{\kappa} \rightarrow \frac{\mu^2 + \beta^2}{T} \,,
\end{equation}
which agrees to the Ward identity~\cite{future1}. 

There is no $1/\omega$ pole in Im[$\alpha$] for all cases. For $\bar{\kappa}$,
at $\mu=0$ there is no $1/\omega$ pole in Im[$\bar{\kappa}$]  while in (a) and (b) there is $1/\omega$ pole only in the superconducting phase.  The appearance of this $1/\omega$ pole in $\bar{\kappa}$ may be understood from \eqref{pheno2} as follows. 
In superconducting phase there is a pole in Im[$\sigma$] so Re[$G_{11}(0)$] $\ne 0$. From Figure \ref{results2} we see that  
Re[$G_{12}(0)$] $=$ Re[$G_{11}(0)$]$\mu$ and Re[$G_{12}(0)$]  is finite if $\mu$ is finite. Then, Im[$\bar{\kappa}(0)$] = $\frac{\mu \mathrm{Re}G_{12}(0)}{\omega} \sim \frac{\mu^2 K_s}{\omega} $.  Therefore, if we subtract $\frac{\mu \mathrm{Re}G_{12}(0)}{\omega}$ from Im[$\bar{\kappa}$], $1/\omega$ pole is expected to be disappeared, which we have confirmed numerically.

\section{Conclusions } \label{sec4}

In this paper,  we studied a simple holographic superconductor model incorporating momentum relaxation. 
The model consists of two parts: The HHH model \cite{Hartnoll:2008vx,Hartnoll:2008kx}}  and massless scalar fields sector for momentum relaxation \cite{Andrade:2013gsa}, where the strength of momentum relaxation is parameterised by $\beta$.

One of the interesting features of the model is that the existence of a new type of superconductor induced by $\beta$ at $\mu=0$. While in the HHH model the superconducting phase transition is understood as a competition between $\mu$ and $T$, in this new case, it is a competition between $\beta$ and $T$. The `pairing mechanism' of two cases must be different. In the new type($\beta \ne 0$ and $\mu = 0$), there is no net charge and the mechanism will be due to particle-anti-particle pairs also interacting with $\beta$, which may be interpreted as kind of `impurity'~\cite{Kim:2014bza}.
The electric optical conductivity of this new superconductor satisfies the FGT sum rule too.  
The deficit of the spectral weight may be interpreted as a deficit of particle-anti-particle pairs which are condensed.

With finite $\mu$ and $\beta$ together, two superconducting mechanisms will compete. As a result, 
the dependence of the critical temperature on $\beta/\mu$ is not monotonic: the critical temperature decreases when $\beta/\mu$ is small and increases when $\beta/\mu$ is large. 
It is different from the previous studies.  As momentum relaxation effect increases, in a Q-lattice model~\cite{Ling:2014laa,Andrade:2014xca} and a single scalar model~\cite{Koga:2014hwa}  the critical temperature decreases while in the ionic lattice model~\cite{Horowitz:2013jaa} the critical temperature increases. 
The condensate has the upper bound when $\beta/\mu \rightarrow \infty$ and the lower bound when $\beta/\mu \rightarrow 0$.

We studied optical electric($\sigma$), thermoelectric($\alpha$), and thermal($\bar{\kappa}$) conductivities.  For all three conductivities, at small $\omega$, a two-fluid model with a modified Drude peak works if $\beta/\mu <1$:
\begin{equation} \label{Drude3}
\sigma, \alpha, \kappa \sim i \frac{K_s}{{\omega}} + \frac{K_n \tau}{1-i {\omega} \tau} + K_0 \,,  
\end{equation}
where $K_s$ and $K_n$ are supposed to be proportional to the superfluid density and normal fluid density and $K_0$ is related to pair creation. For $\beta/\mu \ll 1$, $K_0$ becomes negligible. 
The restriction $\beta/\mu <1$ for \eqref{Drude3} is consistent with the result in metal phase, where coherent metal becomes incoherent metal when $\beta/\mu > 1$ and the Drude peak does not work~\cite{Kim:2014bza}. 
However, the Ferrell-Glover-Tinkham (FGT) sum rule is satisfied for all cases regardless of $\beta/\mu$. 

We have not fully analyzed the parameters, $K_s$, $K_n$, $K_0$, $\tau$, their physical meanings and relations. 
The temperature dependence of the parameters are of physical importance. For example, $K_n$ may be related to 
the energy gap as studied in \cite{Horowitz:2013jaa,Zeng:2014uoa,Ling:2014laa,Andrade:2014xca}. The $\beta$ dependence of $\tau$ is
relevant to the nature of dissipation.  The correct identification of the superfluid density, proportional to $K_s$,  will be essential to investigate Homes' law \cite{Homes:2004wv} holographically \cite{Erdmenger:2012ik}. It will be also useful to  obtain analytic formula for DC conductivities from the horizon data in superconducting phase as in metallic phase~\cite{Donos:2014cya}. 
While the model we considered shows many interesting features as metal and superconductor, it also has shortcomings. 
The electric DC conductivity in normal phase is temperature independent and the insulator phase is lacking. 
It would be interesting to consider superconducting phase without those shortcomings. 
Indeed, there is a simple generalization of the model that provides a temperature dependent DC conductivity 
and an insulating phase at small temperature~\cite{Baggioli:2014roa}, so it would be interesting to construct 
a superconductor model based on this background. It would be also interesting to extend our model to 
the d-wave superconductors~{\cite{Benini:2010qc, Kim:2013oba} and to consider dynamical gauge fields~\cite{Domenech:2010nf}.

\acknowledgments

We would like to thank Kiseok Kim, Dongsu Park, Yunseok Seo and Sang-Jin Sin for valuable discussions and correspondence.
The work of KYK and KKK was supported by Basic Science Research Program through the National Research Foundation of Korea(NRF) funded by the Ministry of Science, ICT \& Future Planning(NRF-2014R1A1A1003220). 
M. Park was supported by by the National Research Foundation of Korea (NRF) funded
by the Korea government with the grant No. 2013R1A6A3A01065975 and TJ Park Science Fellowship of POSCO TJ Park Foundation.
We acknowledge the hospitality at APCTP(``Aspects of Holography'', Jul. 2014) and Orthodox Academy of Crete(``Quantum field theory, string theory and condensed matter physics'', Sep. 2014), where part of this work was done. 

\appendix

\section{One point functions} \label{apa}
%
We briefly summarize how to compute one point functions holographically. Let us consider an ADM decomposition  as follows:
\begin{align}
\dd s^2 =\gamma_{\mu\nu} \dd x^\mu \dd x^\nu  + N^2 \dd r^2 ~~,
\end{align}
where, in our case, the shift vectors vanish and $N$ is the lapse function given by $\frac{1}{\sqrt{\mathcal G (r)}}$. 
The outward pointing normal vector is $n^M =(0,0,0,  \sqrt{\mathcal G (r)} )$ and the extrinsic curvature is given as $K_{\mu\nu}=- \frac{1}{2N} \gamma'_{\mu\nu}$.
 
Let us consider a renormalised action($S_{\mathrm{ren}} $) consisting of a bare action($S_0$) and a counter action($S_{\mathrm{ct}}$) by taking into account  the holographic renormalization:
\begin{equation}
S_{\mathrm{ren}} = S_0+ S_{\mathrm{ct}}~~,
\end{equation}
where 
\begin{equation}
\begin{split}
 S_0&= S_{\mathrm{HHH}} + S_\psi+S_{\mathrm{GH}} \\
 &=\frac{1}{16\pi G}\int_\mathcal{M}  \dd^{4} x \sqrt{-g}\left(   R^{(3)} -K_{\mu\nu}K^{\mu\nu} + K^2  +\frac{6}{L^2}  \right.\\
  &\qquad \qquad \qquad \qquad \qquad \left. -\frac{1}{4}F^2 - \frac{1}{2} \sum_{I=1}^2 (\partial \psi_I)^2   -| D \Phi  |^2  -m^2 |\Phi|^2   \right) \,,
\end{split}
\end{equation}
and 
\begin{equation}
\begin{split}
S_{\mathrm{ct}} = \frac{1}{16\pi G} \int_{\partial \mathcal{M}}  \dd^3 x  \sqrt{-\gamma}~ \left(~   -\frac{4}{L}  - R[\gamma] +\frac{L}{2} \nabla \psi_I \cdot \nabla   \psi_I  \right.  \\
 +  \eta_1   ( \Phi^*  n^A  \partial_A \Phi+\Phi  n^A  \partial_A \Phi^*  )     +\eta_2  \Phi^* \Phi /L
 \Big)~~.
\end{split}
\end{equation}
To fix $\Phi^{(1)}$ on the boundary  we choose $(\eta_1, \eta_2) =(0,-1)$ while to fix $\Phi^{(2)}$ we choose $(\eta_1, \eta_2) =(1,1)$~ \cite{Hartnoll:2008kx}. Variations of the on-shell action with respect to fields yield
\begin{align}
 &\Pi^{\mu\nu} =   \frac{\delta S_{\mathrm{ren}}}{\delta \gamma_{\mu\nu}}|=  \sqrt{-\gamma}  \left(K^{\mu\nu}  -\gamma^{\mu\nu} K  \right)+ \frac{\delta S_{\mathrm{ct}}}{\delta \gamma_{\mu\nu}} \,, \\
 &\Pi^\mu =   \frac{\delta S_{\mathrm{ren}}}{\delta A_{\mu}}|= -\sqrt{-g}F^{r\mu} \,, \\
 &\Pi^I  =   \frac{\delta S_{\mathrm{ren}}}{\delta \psi_{I} }|= -\sqrt{-g} \nabla^r \psi_I  +  \frac{\delta S_{\mathrm{ct}}}{\delta \psi_I}~~,
\end{align}
where  the variations from the counter action are 
\begin{align}
\begin{split}
\frac{\delta S_{ct}}{\delta \gamma_{\mu\nu}}
 =&\sqrt{-\gamma}\left(  - \frac{2}{L}\gamma ^{\mu \nu } + G^{\mu\nu}[\gamma]- \frac{L}{2} \nabla^{\mu }\psi _I \nabla^{\nu }\psi _I +\frac{L}{4} \gamma ^{\mu \nu }\nabla \psi _I \cdot \nabla\psi _I \right.\\ &\left.~~~~~~~~ - \frac{1}{2} \gamma^{\mu\nu}(2\eta_1|\Phi|  n^A  \partial_A |\Phi|    +\eta_2  |\Phi|^2/L )  \right) \,,
 \end{split} \\
\frac{\delta S_{ct}}{\delta \psi_I} =& -\sqrt{-\gamma} L\square_\gamma \psi_I~~.
\end{align}
The expectation values of the energy momentum tensor, the current and the scalar operators in the dual field theory can be computed as 
\begin{equation}\label{boundary tensors}
\begin{split}
&\left< T_{\mu\nu} \right>=  \lim_{r \to \infty} r  \frac{2}{\sqrt{-\gamma}}\Pi_{\mu\nu} \,,  
\quad \left<J^\mu \right>  =  \lim_{r \to \infty} r^3  \frac{1}{\sqrt{-\gamma}}\Pi^{\mu} \,,
\quad \left<\mathcal{O}_I \right> = \lim_{r \to \infty} r^3  \frac{1}{\sqrt{-\gamma}}\Pi^{I} \,, \\
&\left< \mathcal{O}_\Phi^{\Delta =2}  \right> =  \lim_{r \to \infty} r^{-1} \left(  -\sqrt{-g} \nabla^r \Phi -  \sqrt{-\gamma}~ \Phi/L  \right) \,,\quad 
\left< \mathcal{O}_\Phi^{\Delta =1}  \right> =  \lim_{r \to \infty} r^{-2} \left(  \sqrt{-\gamma}~ \Phi/L  \right) \,.
\end{split}
\end{equation}


\providecommand{\href}[2]{#2}\begingroup\raggedright\endgroup

\end{document}